\documentclass[12pt]{article}
\usepackage{subfigure}
\usepackage{latexsym}
\usepackage{amsmath}
\usepackage{amssymb}
\usepackage{amsthm}
\usepackage{undertilde}
\usepackage{graphicx}
\usepackage{hyperref}
\usepackage[latin1]{inputenc}
\usepackage[T1]{fontenc}
\usepackage[all]{xy}
\usepackage{lscape}
\makeatletter
\renewcommand{\@seccntformat}[1]
{\csname the#1\endcsname.\enspace} \makeatother
\def \XXint#1#2#3{{\setbox0=\hbox{$#1{#2#3}{\int}$}
\vcenter{\hbox{$#2#3$}}\kern-.5\wd0}}

\newtheorem{theorem}{Theorem}

\newtheorem{remark}{Remark}

\newtheorem{corollary}{Corollary}

\begin{document}
\begin{center}
{\bf A k-Inflated
Negative Binomial Mixture Regression Model: Application to Rate--Making Systems}\\
\today\\
{\sc Amir T. Payandeh Najafabadi$^{a,}$\footnote{Corresponding
author: amirtpayandeh@sbu.ac.ir} \& Saeed MohammadPour$^b$}
\end{center}
a Department of Mathematical Sciences,
Shahid Beheshti University, G.C. Evin, 1983963113, Tehran, Iran.\\
b  E.C.O College of Insurance, Allameh Tabatab\'ai University,
Tehran, Iran.
\begin{center}
\noindent {\bf Abstract}\\
\end{center}
This article introduces a k-Inflated Negative Binomial mixture
distribution/regression model as a more flexible alternative to
zero-inflated Poisson distribution/regression model. An EM
algorithm has been employed to estimate the model's parameters.
Then, such new model along with a Pareto mixture model have been
employed to design an optimal rate--making system. Namely, this
article employs number/size of reported claims of Iranian third
party insurance dataset. Then, it employs the k-Inflated Negative
Binomial mixture distribution/regression model as well as other
well developed counting models along with a Pareto mixture model
to model frequency/severity of reported claims in Iranian third
party insurance dataset. Such numerical illustration shows that:
({\bf 1}) the k-Inflated Negative Binomial mixture models provide
more fair rate/pure premiums for policyholders under a
rate--making system; and ({\bf 2}) in the situation that number of
reported claims uniformly distributed in past experience of a
policyholder (for instance $k_1=1$ and $k_2=1$ instead of $k_1=0$
and $k_2=2$). The rate/pure premium under the k-Inflated Negative
Binomial mixture models are more appealing
and acceptable.\\


\section{Introduction}
Modeling count data is an interesting topic in variety fields of
applied sciences, such as actuarial sciences, economics,
sociology, engineering, etc. In many practical situation the
popular classical Poisson regression model fails to model count
data which exhibit overdispersion (i.e., the variance of the
response variable exceeds its mean). Moreover, strict assumptions
on Poisson distribution make it more less applicable in situation
that such assumption cannot be strictly verified. The Negative
Binomial distribution/regression has become more and more popular
as a more flexible alternative to Poisson distribution/regression.
In a situation that strict requirements for Poisson distribution
cannot be verified, the Negative Binomial distribution is an
appropriate choice (Johnson et al., 2005). Moreover, the Negative
Binomial is an appropriate choice for overdispersed count data
that are not necessarily heavy--tailed (Aryuyuen \& Bodhisuwan,
2013).

For count data, the overdispersed behavior has been arrived by
{\it either} observing excess of a single value more than number
of expected under the model {\it or} the target population
consisting of several sub-populations. Using k-Inflated and
mixture models are two popular statistical approach to dealing
with an overdispersed behavior. Simar (1976) and Laird (1978) were
two authors who employed Poisson mixture models to considering an
overdispersed behavior. Lambert (1992) considered a zero-inflated
Poisson regression model to take into account an overdispersed
behavior. Wedel et al. (1993) Br\"anna\"as \& Rosenqvist (1994),
Wang et al. (1996), Alf\'o \& Trovato (2004), Wang et al. (1998),
among others, developed idea of using a finite mixture Poisson
regression model to handel overdispersion.

Greene (1994) and  Hall (2000) were pioneer authors who employed
zero-inflated Negative Binomial regression to model
overdispersion. The ordinary Negative Binomial distribution can be
viewed as a mixture of Poisson and gamma distributions (Simon,
1961). To handel an overdispersion phenomena, several extension of
Negative Binomial distribution have been introduced by authors.
For instance Negative Binomial exponential distribution (Panjer \&
Willmot, 1981), Negative Binomial Pareto distribution (Meng et
al., 1999), Negative Binomial Inverse Gaussian distribution
(G\'omez-D\'eniz et al., 2008), Negative Binomial Lindley
distribution (Zamani  \& Ismail, 2010), Negative Binomial Beta
Exponential distribution (Pudprommarat, 2012), and Negative
Binomial Generalized Exponential distribution (Pudprommarat et
al., 2012).

In 2014, Lim et al. considered a k-Inflated Poisson mixture model
which simultaneously takes into account both inflated and mixture
approaches to handel an overdispersion phenomena. Moreover Tzougas
et al. (2014)' introduced a Negative Binomial mixture model to
model an overdispersion phenomena. This article follows Lim et al.
(2014)'s and generalized Tzougas et al. (2014)'s findings. More
precisely, It introduces a k-Inflated Negative Binomial mixture
distribution/regression. To show practical application of our
finding, we consider the problem of designing an optimal
rate--making system. Then, premium of such optimal rate--making
system has been evaluated using the result of this article.

This article has been structured as follows. The k-Inflated
Negative Binomial mixture model, some of its properties, and an EM
algorithm, to estimated its parameters, have been developed in
Section 2. The Pareto mixture regression model has been given in
Section 3. Application of the k-Inflated Negative Binomial mixture
model along with a Pareto mixture model to design an optimal
rate--making system have been given in Section 4. Section 5
employs our model, as well as other well-known model, to evaluate
rate, base, and pure premiums under a rate--making system for
Iranian third party insurance dataset. Base upon three comparison
methods, Section 6 shows that our model provides more accurate (in
some sense) results. Section 7 employs our well fitted models to
calculate rate and pure premiums under two different Scenarios.
Conclusion and suggestions have been given in Section 8

\section{k-Inflated Negative Binomial mixture regression model}
The k-Inflated Negative Binomial mixture, say kINBM, distribution
arrives by combining $m$ weighted mixture Negative Binomial
distribution with a single mass at point $k.$ The probability mass
function for a kINBM distribution has been given by \footnotesize{
\begin{eqnarray}
\label{kINBM_distribution} P(Y=y|{\boldsymbol \theta}) &=&
  p_1I_{\left(y=k\right)}+\sum_{j=2}^{m}p_j\left(\genfrac{}{}{0pt}{}{y+{\alpha }_j-1}{y}\right){\left(\frac{{\tau }_j}{{\alpha }_j+{\tau }_j}\right)}^{{\alpha }_j}{\left(\frac{{\alpha }_j}{{\alpha }_j+{\tau
  }_j}\right)}^yI_{{\Bbb
N}}(y),
\end{eqnarray}}\normalsize
where $k\in{\Bbb N}$ and ${\boldsymbol\theta}$ stands for all $3m$
unknown parameters. Moreover, $\sum_{j=1}^{m}p_j=1$ and
$p_j,\alpha_j,\tau_j\geq0,$ for all $j=1,\cdots,m.$ By a
straightforward calculation, one may show that
\begin{eqnarray*}
E(Y) &=& p_1k+\sum_{j=2}^{m}p_j\alpha_j^2/\tau_j \\
M_Y(t) &=& p_1e^{kt}+ \sum_{j=2}^{m}p_j\left(\frac{\tau_j}{\tau_j+\alpha_j(1-e^t)} \right)^{\alpha_j}, t\leq-\max\left\{log(\frac{\alpha_j}{\alpha_j+\tau_j}),j=2,\cdots,m\right\}\\
F_Y(r)
&=&p_1I_{[k,\infty)}(r)+1-p_1+\sum_{j=2}^{m}p_jRIBeta_{\alpha_j/(\alpha_j+\tau_j)}(r+1;\alpha_j),
\end{eqnarray*}
where $RIBeta_x(a;b)=\int_0^x
t^{a-1}(1-t)^{b-1}dt\Gamma(a+b)/(\Gamma(a)\Gamma(b)),$ for
$a,b\geq0, x\in[0,1],$ stands for the regularized incomplete beta
function.

It is well-known that a Negative Binomial distribution can be
arrived by mixing two Poisson and gamma distributions (Simon,
1961). The following generalized the above fact to the kINBM
distribution.
\begin{corollary}\label{kINBM_Corollary}
Suppose random variable $Y,$ given parameter $\Lambda=\lambda,$
has been distributed according to a k-Inflated Poisson
distribution with probability mass function
$P(Y=y|\lambda)=pI_{\{k\}}(y)+q\exp(-\lambda_i)(\lambda_i)^y/y!,$
where $p~\&~q\in[0,1]$ and $p+q=1.$ Moreover, suppose that
parameter $\lambda$ has been distributed according to a finite
mixture gamma distribution
$f_{\Lambda}(\lambda)=\sum^m_{j=1}{{\varphi }_j{\lambda }^{{\alpha
}_j-1}{{\tau }_j}^{{\alpha }_j}e^{-{\tau }_j\lambda
}/\mathit{\Gamma}({\alpha }_j)},$ where, for all $j=1,\cdots,m,$
${\varphi}_j\in[0,1]$ and $\sum^m_{j=1}{{\varphi }_j}=1.$ Then,
unconditional distribution of $Y$ has a kINB finite mixture
distribution with probability mass function
\begin{equation}\label{EQ_kINBM_Corollary}
P\left(Y=y\right)=pI_{\{k\}}(y)+q\sum^m_{j=1}{{\varphi
}_j\left(\genfrac{}{}{0pt}{}{y+{\alpha
}_j-1}{y}\right){(\frac{{\tau }_j}{1+{\tau }_j})}^{{\alpha
}_j}{(\frac{1}{1+{\tau }_j})}^y}.
\end{equation}
\end{corollary}
For practical application, in Equation \eqref{EQ_kINBM_Corollary},
we set $q:=\sum_{s=1}^{m}\omega_s$ and
${\varphi}_j:=\omega_j/\sum_{s=1}^{m}\omega_s.$

Now, to formulated a kINBM regression model, suppose that for an
$i^{\hbox{th}}$ individual, information on count response
variables $Y_{i1},\cdots, Y_{it}$ along with information on $p$
covariates $X_1,\cdots,X_p$ are available. Also suppose that
$Y_{il}$ given parameter $\Lambda_{il}=\lambda_{il}$ has been
distributed according to a k-Inflated Poisson distribution with
probability mass function
$P(Y_{il}=y_{il}|\lambda_{il})=pI_{\{k\}}(y_{il})+q\exp(-\lambda_{il})(\lambda_{il})^y_{il}/y_{il}!,$
where $p~\&~q\in[0,1]$ and $p+q=1.$ Moreover, suppose that
parameter $\lambda_{il}$ can be evaluated by the following
regression model
\begin{equation*}
{\mathrm{log} \left({\lambda}_{il}\right)=\
}\beta_{0il}+\sum_{k=1}^{p}\beta_{kil}x_{ki}+\epsilon_i,
\end{equation*}
where $\beta_{0i},\cdots,\beta_{pi}$ are regression coefficients
and $u_i=\exp(\epsilon_i)$ has been distributed according to a
finite mixture gamma distribution with density function
\begin{equation}\label{density_finite_mixture_gamma}
f_{U_i}(u_i)=\sum^m_{j=2}{{\varphi }_j\frac{{u_i}^{{\alpha
}_j-1}{{\alpha
}_j}^{{\alpha}_j}e^{-{\alpha}_ju_i}}{\mathit{\Gamma}({\alpha
}_j)}},
\end{equation}
where $\sum^m_{j=2}{\varphi }_j=1$ and ${\alpha}_j\geq0$. To have
$E(\epsilon_i)=0,$ we set both parameters of all gamma
distributions, in the finite mixture gamma distribution, to be
equal.

Using the law of total probability and setting
$d_{il}:=\beta_{0il}+\sum_{k=1}^{p}\beta_{kil}x_{ki},$ one may
show that
\begin{eqnarray*}
  P(Y_{il}=y_{il}|\theta) &=&
  \int_{0}^{\infty}P(Y_{il}=y_{il}|\theta,u_i)f_{U_i}(u_i)du_i\\
  &=&pI_{\{k\}}(y_{il})+\sum^m_{j=2}q{\varphi }_j\dfrac{d_{ijl}^{y_{il}}\alpha_j^{\alpha_j}}{y_{il}!\Gamma(\alpha_j)}\int_{0}^{\infty}e^{-(d_{ijl}+\alpha_j)u_i}u_i^{y_{il}+\alpha_j-1} du_i\\
  &=&pI_{\{k\}}(y_{il})+\sum^m_{j=2}q{\varphi
  }_j\frac{\Gamma(y_{il}+\alpha_j)}{y_{il}!\Gamma(\alpha_j)}\dfrac{d_{ijl}^{y_{il}}\alpha_j^{\alpha_j}}{(d_{ijl}+\alpha_j)^{y_{il}+\alpha_j}},
\end{eqnarray*}
where $\theta$ stands for all unknown parameters. Now by setting
$p=1/(1+\sum^m_{s=2}{e^{\omega_s}}),$
$q{\varphi}_j=e^{\omega_j}/(1+\sum^m_{s=2}{e^{\omega_s}}),$ for
$j=2,\dots,m,$ and
$\left(\genfrac{}{}{0pt}{}{y_{il}+{\alpha}_j-1}{y_{il}}\right):=\frac{\Gamma(y_{il}+\alpha_j)}{y_{il}!\Gamma(\alpha_j)},$
the kINBM regression model can be restated as
\begin{eqnarray}
\label{kINBM_Regression}  \displaystyle
P\left(Y_{il}=y_{il}\right)
&=&\frac{1}{1+\displaystyle\sum^m_{j=2}{e^{\omega_j}}}I_{\{k\}}(y_{il})\\
\nonumber
 &&+\sum^m_{j=2}{\frac{e^{\omega_j}}{1+\displaystyle\sum^m_{s=2}{e^{\omega_s}}}\left(\genfrac{}{}{0pt}{}{y_{il}+{\alpha}_j-1}{y_{il}}\right){t_{ilj}}^{{\alpha}_j}{\left(1-t_{ilj}\right)}^{y_{il}}},,
\end{eqnarray}
where $t_{ilj}:=\alpha_j/(\alpha_j+\exp\{X_iB_{il}\})$ and
$X_iB_{il}:=\beta_{0il}+\sum_{k=1}^{p}\beta_{kil}x_{ik}.$

\subsection*{Parameters estimation}
All unknown parameters of the kINBM regression
\eqref{kINBM_Regression} can be represented as
${\boldsymbol\theta}:=({\boldsymbol\omega},{\boldsymbol\alpha},
{\boldsymbol B}).$ Now to provide a Maximum likelihood estimator,
say MLE, for ${\boldsymbol\theta},$ one may employ an EM
algorithm. In statistical literature, the EM algorithm is a
well-known and practical method to obtain the Maximum likelihood
estimators for parameters in an arbitrary finite mixture model
(McLachlan \& Krishnan, 1997). Now suppose that number of
components, $m,$ is given, and ${\boldsymbol{\
}\boldsymbol{v}}_i=(v_{i1},\cdots,v_{im})$ stands for the latent
vector of component indicator variables, where for $i=1,\cdots,n$
and $j=1,\cdots,m,$ $v_{ij}=1$ whenever observation $i$ comes from
$j^{\hbox{th}}$ component and $v_{ij}=0,$ otherwise. Therefore, we
assume that each observation has been arrived from one of the $m$
components, but the component it belongs to is unobservable and
therefore considered to be the missing data.

Now using the Multinomial distribution for the unobservable vector
${\boldsymbol{\ }\boldsymbol{v}}_i,$ the complete data
loglikelihood function, for, the kINB regression model, can be
written as the following, see Rigby \& Stasinopoulos (2009) for an
update information.\footnotesize{
\begin{eqnarray}\label{complete_loglikelihood}
  l_c\left({\boldsymbol\theta}|y_i,{\boldsymbol
v}_i,X_i\right) &=& \sum^n_{i=1}v_{i1}\mathrm{log}
\left(\frac{1}{1+\sum^m_{l=2}\exp\left(\omega_l\right)}\right)I_{\left(y_i=k\right)}\\
\nonumber&&+\sum^n_{i=1}\sum^m_{j=2}v_{ij}\mathrm{log}
\left(\frac{\exp\left(\omega_j\right)}{1+\sum^m_{l=2}{\exp\left(\omega_l\right)}}\left(\genfrac{}{}{0pt}{}{y_i+\alpha_{j}-1}{y_i}\right){t_j}^{\alpha_j}{\left(1-t_j\right)}^{y_i}\right),
\end{eqnarray}}\normalsize
where
${\boldsymbol\theta}:=({\boldsymbol\omega},{\boldsymbol\alpha},
{\boldsymbol B})$ stands for all unknown parameters,
$t_j:=\alpha_j/(\alpha_j+\exp\{X_iB_j\}),$ and
$X_iB_j:=\beta_{0j}+\sum_{k=1}^{p}\beta_{kj}x_{ik}.$

The EM algorithm employs the following two steps to maximize the
above loglikelihood function.
\begin{description}
    \item[E-step:] In this step, using given data along with current estimates ${\widehat{\boldsymbol\theta}}^{(r)}:=({\widehat{\boldsymbol\omega}}^{(r)},{\widehat{\boldsymbol\alpha}}^{(r)},
{\widehat{\boldsymbol B}}^{(r)})$ obtained from the
$r^{\hbox{th}}$ iteration, the probability ${\hat{v}}_{ij}$
estimates. This probability, $(r+1)^{\hbox{th}}$ iteration, can be
stated as\footnotesize{
\begin{eqnarray*}
  \hat{v}_{ij}^{(r+1)} &=&
  E\left[v_{ij}|y_i,X_i,{\widehat{\boldsymbol\theta}}^{(r)}\right]=1\times
  P\left(v_{ij}=1|y_i,X_i,{\widehat{\boldsymbol\theta}}^{(r)}\right)+0\times
  P\left(v_{ij}=0|y_i,X_i,{\widehat{\boldsymbol\theta}}^{(r)}\right)\\
  &=&\frac{f\left(y_i|v_{in}=1,X_i,{\widehat{\boldsymbol\theta}}^{(r)}\right)P\left(v_{in}=1|X_i,{\widehat{\boldsymbol\theta}}^{(r)}\right)}{f\left(y_i|X_i,{\widehat{\boldsymbol\theta}}^{(r)}\right)}\\
  &=&\frac{\exp\left({\hat{\omega}}^{\left(r\right)}_j\right)\left(\genfrac{}{}{0pt}{}{y_i+{\widehat{\alpha
}}^{\left(r\right)}_j-1}{y_i}\right){\left({\widehat{\boldsymbol
t}}^{(r)}_j\right)}^{{\widehat{\alpha
}}^{\left(r\right)}_j}{\left(1-{\widehat{\boldsymbol
t}}^{(r)}_j\right)}^{y_i}}{1+\sum^m_{s=2}{\exp\left({\hat{\omega}}^{\left(r\right)}_s\right)\left(\genfrac{}{}{0pt}{}{y_i+{\widehat{\alpha
}}^{\left(r\right)}_s-1}{y_i}\right){\left({\widehat{\boldsymbol
t}}^{(r)}_s\right)}^{{\widehat{\alpha
}}^{\left(r\right)}_s}{\left(1-{\widehat{\boldsymbol
t}}^{(r)}_s\right)}^{y_i}}},
\end{eqnarray*}}\normalsize
where ${\widehat{\boldsymbol
t}}^{(r)}_j:={\widehat\alpha}^{(r)}_j/({\widehat\alpha}^{(r)}_j+\exp\{X_i{\widehat
B}^{(r)}_j\}),$ and $X_i{\widehat
B}^{(r)}_j:={\widehat\beta}^{(r)}_{0j}+\sum_{k=1}^{p}{\widehat\beta}^{(r)}_{kj}x_{ik}.$
    \item[M-step:] Given the probability $\hat{v}_{ij},$ this step
    maximizes, in the $(r+1)^{\hbox{th}}$ iteration, the following loglikelihood $Q(\cdot)$ with respect
    to ${\boldsymbol\theta}:=({\boldsymbol\omega},{\boldsymbol\alpha},
{\boldsymbol B}).$\footnotesize{
\begin{eqnarray*}
  \boldsymbol{Q} &=&
  E\left[l_c|y_i,X_i,{\widehat{\boldsymbol\theta}}^{(r)}\right]\\
  &=& \sum^n_{i=1}{{{\hat{v}}_{i1}}^{\left(r+1\right)}{\mathrm{\log} \left(\frac{1}{1+\sum^m_{l=2}{\exp\left(\omega_l\right)}}\right)\ }I_{\left(y_i=k\right)}+\sum^n_{i=1}\sum^m_{l=2}{\left[{{\hat{v}}_{il}}^{\left(r+1\right)}{\mathrm{log} \left(\frac{\exp\left({\widehat{\boldsymbol \omega}}^{(r)}_l\right)}{1+\sum^m_{s=2}{\exp\left({\widehat{\boldsymbol \omega}}^{(r)}_s\right)}}\right)\ }\right]}}\\
  &&+ \sum^n_{i=1}{\sum^m_{l=2}{{{\hat{v}}_{il}}^{\left(r+1\right)}{\mathrm{\log} \left(\left(\genfrac{}{}{0pt}{}{y_i+{{\widehat{\boldsymbol \alpha}}^{(r)}}_l-1}{y_i}\right){\left(\frac{{{\widehat{\boldsymbol \alpha}}^{(r)} }_l}{{{\widehat{\boldsymbol \alpha}}^{(r)} }_l+\exp(X_i{\widehat{\boldsymbol B}}^{(r)}_l)}\right)}^{{{\widehat{\boldsymbol \alpha}}^{(r)} }_l}{\left(\frac{\exp(X_i{\widehat{\boldsymbol B}}^{(r)}_l)}{{{\widehat{\boldsymbol \alpha}}^{(r)} }_l+\exp(X_i{\widehat{\boldsymbol B}}^{(r)}_l)}\right)}^{y_i}\right)\
  }}}\\
  &=:& Q_1+Q_2.
\end{eqnarray*}}\normalsize
Updated parameters
${\widehat{\boldsymbol\theta}}^{(r+1)}:=({\widehat{\boldsymbol\omega}}^{(r+1)},{\widehat{\boldsymbol\alpha}}^{(r+1)},
{\widehat{\boldsymbol B}}^{(r+1)})$ have been arrived by solving
the following equation.\footnotesize{
\begin{eqnarray*}
  \frac{\partial Q_1}{\partial \omega_j} &=& \sum^n_{i=1}{\frac{\partial }{\partial\omega_j}{{\hat{v}}_{i1}}^{\left(r+1\right)}{\mathrm{log}
\left(\frac{1}{1+\sum^m_{l=2}{\exp\left(\omega_l\right)}}\right)\
}I_{\left(y_i=k\right)}+\sum^n_{i=1}\frac{\partial
}{\partial\omega_j}{{\hat{v}}_{ij}}^{\left(r+1\right)}{\mathrm{log}
\left(\frac{\exp\left(\omega_j\right)}{1+\sum^m_{l=2}{\exp\left(\omega_l\right)}}\right)\
}}=0; \\
  \frac{\partial Q_2}{\partial B_j} &=& \sum^n_{i=1}\frac{\partial}{\partial
B_j}{{\hat{v}}_{in}}^{\left(r+1\right)}{\mathrm{log}
\left(\left(\genfrac{}{}{0pt}{}{y_i+{\alpha
}_j-1}{y_i}\right){\left(\frac{{\alpha }_j}{{\alpha
}_j+exp(X_iB_j)}\right)}^{{\alpha
}_j}{\left(\frac{\exp(X_iB_j)}{{\alpha
}_j+exp(X_iB_j)}\right)}^{y_i}\right)\ }=0; \\
  \frac{\partial Q_2}{\partial {\alpha}_j} &=& \sum^n_{i=1}\frac{\partial
}{\partial
\alpha_j}{{\hat{v}}_{in}}^{\left(r+1\right)}{\mathrm{log}
\left(\left(\genfrac{}{}{0pt}{}{y_i+{\alpha
}_j-1}{y_i}\right){\left(\frac{{\alpha }_j}{{\alpha
}_j+exp(X_iB_j)}\right)}^{{\alpha
}_j}{\left(\frac{\exp(X_iB_j)}{{\alpha
}_j+exp(X_iB_j)}\right)}^{y_i}\right)\ }=0.
\end{eqnarray*}}\normalsize
Since the above three equations cannot solve explicitly, such
updated parameters have been obtained using the following
Iteratively Reweighted Least Squares, say IRLS, method.
\begin{eqnarray*}
  {\hat{\omega}}^{\left(r+1\right)}_j &=& {\hat{\omega}}^{\left(r\right)}_j+{\left(E\left(\frac{-{\partial }^2Q_1}{\partial
{\omega_j}^2}\right)\right)}^{-1}.\frac{\partial Q_1}{\partial \omega_j}; \\
  {\hat{B}}^{\left(r+1\right)}_j &=& {\hat{B}}^{\left(r\right)}_j+{\left(E\left(\frac{-{\partial }^2Q_2}{\partial
{B_j}^2}\right)\right)}^{-1}.\frac{\partial Q_2}{\partial B_j};\\
  {\widehat{\alpha }}^{\left(r+1\right)}_j &=& {\widehat{\alpha
  }}^{\left(r\right)}_j+{\left(E\left(\frac{-{\partial }^2Q_2}{\partial {{\alpha
}_j}^2}\right)\right)}^{-1}.\frac{\partial Q_2}{\partial {\alpha
}_j}.
\end{eqnarray*}
In IRLS method, $E\left(-{\partial}^2\left({\mathrm{log}
\mathrm{likelihood}\ }\right)/\partial
{\mathrm{parameter}}^{\mathrm{2}}\right)$ can be viewed as the
Fisher information matrix and $\partial \left({\mathrm{log}
\mathrm{likelihood}\ }\right)/\partial \mathrm{parameter}$ as
score function.
\end{description}
After updated Parameter estimates
${\widehat{\boldsymbol\theta}}^{(r+1)}:=({\widehat{\boldsymbol\omega}}^{(r+1)},{\widehat{\boldsymbol\alpha}}^{(r+1)},
{\widehat{\boldsymbol B}}^{(r+1)}),$ the complete data
loglikelihood for $(r+1)^{\hbox{th}}$ iteration, arrives
by\footnotesize{
\begin{eqnarray*}
  l^{\left(r+1\right)}_c&=&\sum^n_{i=1}{{{\hat{v}}_{i1}}^{\left(r+1\right)}{\mathrm{log} \left(\frac{1}{1+\sum^m_{l=2}{\exp\left({\hat{\omega}}^{\left(r+1\right)}_l\right)}}\right)\
  }I_{\left(y_i=k\right)}}\\
  &&+\sum^n_{i=1}{\sum^m_{j=2}{{{\hat{v}}_{ij}}^{\left(r+1\right)}{\mathrm{log}
\left(\frac{\exp\left({\hat{\omega}}^{\left(r+1\right)}_j\right)}{1+\sum^m_{l=2}{\exp\left({\hat{\omega}}^{\left(r+1\right)}_l\right)}}\left(\genfrac{}{}{0pt}{}{y_i+{\widehat{\alpha
}}^{\left(r+1\right)}_j-1}{y_i}\right){\left({\widehat{\boldsymbol
t}}^{(r+1)}_j\right)}^{{\widehat{\alpha
}}^{\left(r+1\right)}_j}{\left(1-{\widehat{\boldsymbol
t}}^{(r+1)}_j\right)}^{y_i}\right)\ }}},
\end{eqnarray*}}\normalsize
where ${\widehat{\boldsymbol
t}}^{(r+1)}_j:={\widehat\alpha}^{(r+1)}_j/({\widehat\alpha}^{(r+1)}_j+\exp\{X_i{\widehat
B}^{(r+1)}_j\}),$ and $X_i{\widehat
B}^{(r+1)}_j:={\widehat\beta}^{(r+1)}_{0j}+\sum_{k=1}^{p}{\widehat\beta}^{(r+1)}_{kj}x_{ik}.$
Now, in the E-step $v_{ij}-$s have been estimated. This loop has
been repeated until the difference
$\left|l^{\left(r+1\right)}_c-l^{\left(r\right)}_c\right|$ has
been converged, in some sense.

It is worthwhile to mention that, since regression coefficients
${\boldsymbol B}=(\beta_{0},\cdots,\beta_{p})^\prime$ have been
estimated using the MLE methods. therefore, number of mixture
component impact on such estimators.

\section{Pareto mixture regression model}
The Pareto mixture distribution arrives by combining $m$ weighted
mixture Pareto distributions. The density function for a Pareto
mixture distribution has been given by \footnotesize{
\begin{eqnarray}
\label{Pareto_mixture_distribution} f_{Z}(z|{\boldsymbol
\vartheta}) &=&
\sum_{j=1}^{m}\rho_j\alpha_j\frac{\gamma_j^{\alpha_j}}{(z+\gamma_j)^{\alpha_j+1}},
\end{eqnarray}}\normalsize
where ${\boldsymbol\vartheta}=({\boldsymbol\rho},
{\boldsymbol\alpha}, {\boldsymbol\gamma})$ stands for all $3m$
unknown parameters. Moreover, $\sum_{j=1}^{m}\rho_j=1$ and
$\rho_j,\alpha_j\geq0,$ for all $j=1,\cdots,m.$ More details on
this distribution can be found in Tzougas et al. (2014).

Tzougas et al. (2014) showed that a Pareto mixture distribution
can be arrived by mixing two exponential and inverse gamma
distributions.

Now, to formulated a Pareto mixture regression model, suppose that
for an $i^{\hbox{th}}$  individual, information on continuous
response variables $Z_{i1},\cdots, Z_{it}$ along with information
on $p$ covariates $W_1,\cdots,W_p$ are available. Also suppose
that $Z_{il}$ given parameter $\Theta_{il}=\theta_{il}$ has been
distributed according to an exponential distribution with density
function
$f_{Z_{il}|\Theta_{il}=\theta_{il}}(z_{il})=\exp\{-z_{il}/\theta_{il}
\}/\theta_{il}.$ Moreover, suppose that parameter $\theta_{il}$
can be evaluated by the following regression model
\begin{equation*}
{\mathrm{log} \left({\theta}_{il}\right)=\
}d_{0il}+\sum_{k=1}^{p}d_{kil}w_{ki}+\epsilon_i,
\end{equation*}
where $d_{0i},\cdots,d_{pi}$ are regression coefficients and
$u_i=\exp(\epsilon_i)$ has been distributed according to a finite
mixture Inverse gamma distribution with density function
\begin{equation}\label{density_finite_mixture_Inverse_gamma}
f_{U_i}(u_i)=\sum^m_{j=1}\rho_j\frac{(\alpha_j-1)^{\alpha_j}u_i^{-\alpha_j-1}}{\Gamma(\alpha_j)}e^{-(\alpha_j-1)/u_i},
\end{equation}
where $\sum^m_{j=1}\rho_j=1$ and $\rho_j,~{\alpha}_j\geq0$. To
have $E(\epsilon_i)=0$ in Equation
\eqref{Pareto_mixture_distribution} we set $\gamma_j=\alpha_j-1,$
for $j=1,\cdots,m.$

Using the law of total probability and setting
$b_{il}:=d_{0il}+\sum_{k=1}^{p}d_{kil}w_{ki},$ one may show that
\begin{eqnarray*}
  f_{Z_{il}|\vartheta}(z_{il}) &=&
  \int_{0}^{\infty}f_{Z_{il}||\vartheta,u_i}(z_{il})f_{U_i}(u_i)du_i\\
  &=& \int_{0}^{\infty} e^{-z_{il}\exp\{-b_{il}\}/u_i}\exp\{-b_{il}\}u_i^{-1}\sum^m_{j=1}\rho_j\frac{(\alpha_j-1)^{\alpha_j}u_i^{-\alpha_j-1}}{\Gamma(\alpha_j)}e^{-(\alpha_j-1)/u_i}  du_i\\
  &=&\sum_{j=1}^{m}\rho_j\alpha_j\frac{(\alpha_j-1)^{\alpha_j}}{(z+\alpha_j-1)^{\alpha_j+1}}.
\end{eqnarray*}
Similar to the kINB regression/distribution the maximum liklihood
estimator for parameters of a Pareto mixture
regression/distribution can be obtained using the EM algorithm.
Fortunately,  Rigby \& Stasinopoulos (2001) developed a R package,
named 'GAMLSS`, for such propose, see Rigby \& Stasinopoulos
(2001, 2009) for more details.

\section{Application to posteriori rate--making system}
The rate--making system is a non-life actuarial system which rates
policyholders based upon their last $t$ years record (Payandeh
Najafabadi et al., 2015). A rate--making system based upon
policyholders' characteristics assigns a priori premium for each
policyholder. Then, it employs the last $t$ years claims
experience of each insured to update such priori premium and
provides posteriori premium (Boucher \& Inoussa, 2014). The
Bonus--Malus system is a commercial and practical version of the
rate--making system which takes into account current year
policyholders' experience to determine their next year premium.

There is a considerable attention from authors to study
rate--making systems (or Bonus--Malus systems). For instance:
Several mathematical tools for pricing a rate--making system has
been provided by Lange (1969). Dionne \& Vanasse (1989, 1992)
employed available asymmetric information under Poisson and
Negative Binomial regression models to determine premium of a
rate--making system. In 1995, Lemaire designed an optimal
Bonus--Malus system based on Negative Binomial distribution.
Pinquet (1997) considered Poisson and Lognormal distributions to
design an optimal Bonus--Malus system. Walhin \& Paris (1999)
considered a Hofmann's distribution along with a finite mixture
Poisson distribution to evaluate elements of a Bonus--Malus
system. The relatively premium of a rate--making system under the
exponential loss function has been evaluated by Denuit \& Dhaene
(2001). In 2001, Frangos \& Vrontos designed an optimal
Bonus--Malus system using both Pareto and Negative Binomial
distributions. Using the bivariate Poisson regression model
Berm\'udez \& Morata (2009) studied priori rate--making procedure
for an automobile insurance database which has two different types
of claims. In 2011, Berm\'udez \& Karlis employed a Bayesian
multivariate Poisson model to determine premium of a rate--making
system which has a non-ignorable correlation between types of its
claims. Boucher \& Inoussa (2014) introduced a new model to
determine premium of a rate--making system whenever panel or
longitudinal data are available. The Sichel distribution along
with a Negative Binomial distribution have been considered by
Tzougas \& Frangos (2013, 2014a, 2014b). Tzougas et al. (2014)
employed a finite mixture distribution to model frequency and
severity of accidents. Payandeh Najafabadi et al. (2015) employed
Payandeh Najafabadi (2010)'s idea to determine credibility premium
for a rate--making system whenever number of reported claims
distributed according to a zero-inflated Poisson distribution.
Several authors have been employed zero-inflated models in
actuarial science, see instance Yip \& Yau (2005), Boucher et al.
(2009), Boucher \& Denuit (2008), and Boucher et al. (2007), among
others.

Under a rate--making system the pure premium of an $i^{\hbox{th}}$
policyholder at $(t+1)^{\hbox{th}}$ year has been estimated by
multiplication of estimated base premium, say ${\hat BP}(t+1),$
into corresponding estimated rate premium, say ${\hat Rate}(t+1).$
From decision theory point of view, the Bayes estimator offers an
intellectually and acceptable estimation for both the rate premium
$Rate(t+1)$ and the base premium $BP(t+1).$ Such Bayes estimators,
under the quadratic loss function, can be obtained by posterior
expectation of risk parameters given number and severity of
reported claims at first $t+1$ years, see Denuit et al. (2007) for
more details.

Therefore, to determine premium for $i^{\hbox{th}}$ policyholder,
under a rate--making system, one has to determine both Bayes
estimators. The following two theorems develop such estimators.
Namely, in the first step, it supposes that number of reported
claim $Y_1,\cdots,Y_t,$ given risk parameter
$\Lambda_i=\lambda_i,$ has been distributed according to a
k-Inflated Poisson distribution and risk parameter $\Lambda_i$
distributed as a finite mixture Gamma. In the second step, it
supposes that claim size random variable $Z_1,\cdots,Z_t,$ given
risk parameter $\Theta_i=\theta_i,$ has been distributed according
to an exponential distribution and risk parameter $\Theta_i$
distributed as a finite mixture inverse Gamma. Finally, it derives
such Bayes estimators for risk parameters $\theta_i$ and
$\lambda_i.$

\begin{theorem}
\label{Bayes_estimator_rate_premium} Suppose that for an
$i^{\hbox{th}}$ policyholder, number of reported claims in the
last $t$ years have been restated as ${\bf
Y}_{i}=(Y_{i1},\cdots,Y_{it}).$ Also suppose that, for
$l=1,\cdots,t,$ $Y_{il}$ given parameter
$\Lambda_{il}=\lambda_{ik}$ has been distributed according to a
k-Inflated Poisson distribution with probability mass function
$P(Y_{il}=y_{il}|\lambda_{il})=pI_{\{k\}}(y_{il})+q\exp(-\lambda_{il})(\lambda_{il})^y_{il}/y_{il}!,$
where $p~\&~q\in[0,1]$ and $p+q=1.$ Moreover, suppose that risk
parameter $\Lambda_i$ can be restated as regression model
$\log({\lambda}_{il})=C_{i}B_{il}+{\epsilon}_i$ where ${\bf
C}_{i}=(1,c_{i,1},\dots,c_{i,p})$ is the vector of $p$
characteristics/covariates for an $i^{\hbox{th}}$ policyholder,
$B_{il}=(\beta_{0il},\cdots,\beta_{pil})^\prime$ is the vector of
the regression coefficients, and $u_i=\exp(\epsilon_i)$ has been
distributed according to finite mixture gamma distribution with
density function $f_{U_i}(u_i)=\sum^m_{j=1}{{\varphi
}_j{{u_i}^{{\alpha }_j-1}{{\alpha }_j}^{{\alpha }_j}e^{-{\alpha
}_ju_i}}/{\mathit{\Gamma}({\alpha }_j)}},$ where $u_i>0$, ${\alpha
}_j>0$ and $\sum^m_{j=1}{{\varphi }_j=1}$. Then, Bayes estimator
for the rate premium $\widehat{ Rate}_i(t+1),$ of an
$i^{\hbox{th}}$ policyholder at $(t+1)^{\hbox{th}}$ year, is given
by \footnotesize{
\begin{equation} \label{Eq_Bayes_Rate_Premium} \widehat{ Rate}_i(t+1)=e^{C_{i}B_{it+1}}\frac{\int^{\infty
}_0{u_i\prod^t_{l=1}{h_{il}(u_i)}}\sum^m_{j=1}k_{j}(u_i)d_{u_i}}{\int^{\infty
}_0{\prod^t_{l=1}{h_{il}(u_i)}}\sum^m_{j=1}k_{j}(u_i)d_{u_i}},
\end{equation}}\normalsize
where $k_{j}(u_i):={{\varphi }_j{{u_i}^{{\alpha }_j-1}{{\alpha
}_j}^{{\alpha }_j}e^{-{\alpha }_ju_i}}/{\mathit{\Gamma}({\alpha
}_j)}},$ ${\mathit{\Gamma}(\cdot)}$ stands for the Gamma function,
and
$h_{il}(u_i):=pI_{\left(y_{il}=k\right)}+qe^{-\exp(C_{i}B_{il})u_i}{\left(\exp(C_{i}B_{il})u_i\right)}^{y_{il}}/y_{il}!.$
\end{theorem}
{\it Proof.} The Bayes estimator for the rate premium
$Rate_i(t+1),$ under the quadratic loss function, is mean of
posterior distribution $\Lambda_{it+1}|({\bf Y}_{i}, {\bf
C}_{i}).$ Such the posterior distribution can be restated as the
following.
\begin{eqnarray*}
 f_{\Lambda_{it+1}|({\bf Y}_{i}, {\bf
C}_{i})}(\lambda_{it+1}) &=&
\dfrac{\prod_{l=1}^{t}P(Y_{il}=y_{il}|\Lambda_{il})P(\Lambda_{il}=e^{C_{i}B_{il}}u_i)}{\int_{0}^{\infty}\prod_{l=1}^{t}P(Y_{il}=y_{il}|\Lambda_{il})P(\Lambda_{il}=e^{C_{i}B_{il}}u_i)du_i}\\
&=&\dfrac{\prod^t_{l=1}{h_{il}(u_i)}\sum^m_{j=1}k_{j}(u_i)}{\int_{0}^{\infty}\prod^t_{l=1}{h_{il}(u_i)}\sum^m_{j=1}k_{j}(u_i)d_{u_i}}.
\end{eqnarray*}
Now the desired result arrives by $$\widehat{
Rate}_i(t+1)=\int_{0}^{\infty}e^{C_{i}B_{it+1}}u_if_{\Lambda_{it+1}|({\bf
Y}_{i}, {\bf C}_{i})}(\lambda_{it+1})du_i~~\square$$

In a situation that $q=1,$ the rate premium ${\hat Rate}_i(t+1)$
can be restated as
\begin{eqnarray*}
 \widehat{ Rate}_i(t+1) &=&
 e^{C_{i}B_{it+1}}\dfrac{\sum^m_{j=1}{\varphi}_j\alpha_j^{\alpha_j}\Gamma(\alpha_j+y_{i.}+1)/(\Gamma(\alpha_j)(\alpha_j+\sum_{l=1}^te^{C_{i}B_{il}}))}{\sum^m_{j=1}{\varphi}_j\alpha_j^{\alpha_j}\Gamma(\alpha_j+y_{i.})/(\Gamma(\alpha_j)(\alpha_j+\sum_{l=1}^te^{C_{i}B_{il}}))},
\end{eqnarray*}
where $y_{i\cdot}=\sum_{l=1}^ty_{il}.$ This situation has been
studied by Dionne \&  Vanasse (1992) for an one mixture
distribution and by Tzougas et al. (2014) for an $m$ mixture
distribution. In the case that $t=0,$ one may show that $\widehat{
Rate}_i(1)=e^{C_{i}B_{i1}}.$

\begin{remark}\label{Bayes_estimator_rate_premium-under-distribution}
For the situation that no covariate information has been taken
into account, say a distribution model, and the risk parameter
$\Lambda_i$ has been distributed according to a finite mixture
gamma distribution with density function given by
\eqref{density_finite_mixture_gamma}. Result of Theorem
\eqref{Bayes_estimator_rate_premium} can be reformulated as
\footnotesize{
\begin{equation*} \label{Eq_Bayes_Rate_Premium}
\widehat{ Rate}_i(t+1)=\frac{\int^{\infty
}_0{\lambda_{it+1}\prod^t_{l=1}\left(pI_{\left(y_{il}=k\right)}+q\frac{e^{-\lambda_{it+1}}{\left(\lambda_{it+1}\right)}^{y_{il}}}{y_{il}!}\right)}\sum^m_{j=1}{{\varphi
}_j\frac{{\lambda_{it+1}}^{{\alpha }_j-1}{{\tau}_j}^{{\alpha
}_j}e^{-{\tau}_j\lambda_{it+1}}}{\mathit{\Gamma}({\alpha
}_j)}}d_{\lambda_{it+1}}}{\int^{\infty
}_0{\prod^t_{l=1}\left(pI_{\left(y_{il}=k\right)}+q\frac{e^{-\lambda_{it+1}}{\left(\lambda_{it+1}\right)}^{y_{il}}}{y_{il}!}\right)}\sum^m_{j=1}{{\varphi
}_j\frac{{\lambda_{it+1}}^{{\alpha }_j-1}{{\tau}_j}^{{\alpha
}_j}e^{-{\tau}_j\lambda_{it+1}}}{\mathit{\Gamma}({\alpha
}_j)}}d_{\lambda_{it+1}}}.
\end{equation*}}\normalsize
\end{remark}

The following theorem develops a Bayes estimator for the base
premium $BP_i(t+1)$ for an $i^{\hbox{th}}$ policyholder at
$(t+1)^{\hbox{th}}$ year.

\begin{theorem}\label{Bayes_estimator_base_premium}
Suppose that for an $i^{\hbox{th}}$ policyholder, severity/size of
claims in the last $t$ years have been restated as ${\bf
Z}_{i}=({\bf Z}_{i1},\cdots,{\bf Z}_{it}).$ Also suppose that, for
$l=1,\cdots,t,$ ${\bf Z}_{il}=(Z_{il1},\cdots,Z_{ilk_{il}}),$
where $k_{il}$ stands for number of reported claims by
$i^{\hbox{th}}$ policyholder at $l^{\hbox{th}}$ year, and for
$s=1,\cdots,k_{il},$ assume that $Z_{ils}$ given parameter
$\Theta_{il}=\theta_{il}$ has been distributed according to an
exponential distribution function with density function
$f_{Z_{il}|\Theta_{il}=\theta_{il}}(z_{il})=\exp\{-z_{il}/\theta_{il}\}/\theta_{il}.$
Moreover, suppose that risk parameter $\Theta_i$ can be restated
as $\log({\theta }_{i,l})={\bf W}_{i}{\bf D}_{il}+\epsilon_i,$
where ${\bf W}_{i}=(1,w_{i,1},\dots,w_{i,p})$ is the vector of $p$
characteristics/covariates for an $i^{\hbox{th}}$ policyholder,
${\bf D}_{il}=(d_{0il},\cdots,d_{pil})^\prime$ is the vector of
the regression coefficients, and $u_i=\exp(\epsilon_i)$ has been
distributed according to a finite mixture Inverse Gamma with
density function
$f_{U_i}(u_i)=\sum_{j=1}^{m}\phi_j(\eta_j-1)^{\eta_j}u_i^{-\eta_j-1}\exp(-(\eta_j-1)/u_i)/\Gamma(\eta_j),$
where $u_i>0$, ${\eta}_j>0$ and $\sum^m_{j=1}{\phi_j=1}$. Then,
Bayes estimator for the the base premium $BP_i(t+1)$ for an
$i^{\hbox{th}}$ policyholder at $(t+1)^{\hbox{th}}$ year, is given
by
\begin{equation} \label{Eq_Bayes_Base_Premium}
\widehat{BP}_i^{t+1}=e^{{
W}_{i}{D}_{it+1}}\dfrac{\sum^m_{j=1}\phi_j\frac{(\eta_j-1)^{\eta_j}}{\Gamma(\eta_j)}\frac{\Gamma(\eta_j+K_i-1)}{(\eta_j+\sum_{l=1}^{t}\sum_{s=1}^{k_{il}}\exp(-W_{i}D_{il})z_{ils}-1)^{\eta_j+K_i-1}}}{\sum^m_{j=1}\phi_j\frac{(\eta_j-1)^{\eta_j}}{\Gamma(\eta_j)}\frac{\Gamma(\eta_j+K_i)}{(\eta_j+\sum_{l=1}^{t}\sum_{s=1}^{k_{il}}\exp(-W_{i}D_{il})z_{ils}-1)^{\eta_j+K_i}}},
\end{equation}
where $K_i=\sum_{l=1}^{t}k_{il}.$
\end{theorem}
{\it Proof.}  The posterior distribution of $\Theta_{it+1}|({\bf
Z}_{i}, {\bf W}_{i})$ can be restated as
\begin{eqnarray*}
f_{\Theta_{it+1}|({\bf Z}_{i}, {\bf W}_{i})}(\theta_{it+1}) &=&
  \dfrac{\prod_{l=1}^{t}\prod_{s=1}^{k_{il}}f_{Z_{ils}|\Theta_{il}}(z_{ils})f_{\Theta_{il}}(e^{W_{i}D_{il}}u_i)}{\int_{0}^{\infty}\prod_{l=1}^{t}\prod_{s=1}^{k_{il}}f_{Z_{ils}|\Theta_{il}}(z_{ils})f_{\Theta_{il}}(e^{W_{i}D_{il}}u_i)du_i}\\
  &=&\dfrac{\sum^m_{j=1}\phi_j\frac{(\eta_j-1)^{\eta_j}}{\Gamma(\eta_j)}u_i^{-\eta_j-K_i-1}\exp\{-\frac{\eta_j+\sum_{l=1}^{t}\sum_{s=1}^{k_{il}}\exp(-W_{i}D_{il})z_{ils}-1}{u_i}\}}{\int_{0}^{\infty}\sum^m_{j=1}\phi_j\frac{(\eta_j-1)^{\eta_j}}{\Gamma(\eta_j)}u_i^{-\eta_j-K_i-1}\exp\{-\frac{(\eta_j+\sum_{l=1}^{t}\sum_{s=1}^{k_{il}}\exp(-W_{i}D_{il})z_{ils}-1)}{u_i}\}
  du_i}.
\end{eqnarray*}
The desired Bayes estimator arrives by
$$\widehat{BP}_i^{t+1}=\int_{0}^{\infty}e^{W_{i}D_{it+1}}u_if_{\Theta_{it+1}|({\bf Z}_{i}, {\bf W}_{i})}(\theta_{it+1})du_i.~~\square$$
The above result also obtained by Tzougas et al. (2014).

\begin{remark}\label{Bayes_estimator_base_premium-under-distribution}
For the situation that no covariate information has been taken
into account, say a distribution model, and the risk parameter
$\Theta_{i}$ has been distributed according to a finite mixture
Inverse Gamma with density function given by
\eqref{density_finite_mixture_Inverse_gamma}. Result of Theorem
\eqref{Bayes_estimator_base_premium} can be reformulated as
\begin{equation*}
\widehat{BP}_i^{t+1}=\dfrac{\sum^m_{j=1}\phi_j\frac{(\eta_j-1)^{\eta_j}}{\Gamma(\eta_j)}\frac{\Gamma(\eta_j+K_i-1)}{(\eta_j+\sum_{l=1}^{t}\sum_{s=1}^{k_{il}}z_{ils})^{\eta_j+K_i-1}}}{\sum^m_{j=1}\phi_j\frac{(\eta_j-1)^{\eta_j}}{\Gamma(\eta_j)}\frac{\Gamma(\eta_j+K_i)}{(\eta_j+\sum_{l=1}^{t}\sum_{s=1}^{k_{il}}z_{ils})^{\eta_j+K_i}}}.
\end{equation*}
\end{remark}

To show practical application of our findings, the next section
provides an real example.
\section{Numerical Application}
Now, we considered available data from Iranian third party
liability, at 2011 year. After a primary investigation, we just
trusted information about 8874 policyholders. We used 4
independent variables, as covariates, presented in Table 1. For
each policyholder we have the initial information at the beginning
of the period and we are interested such covariates to model
frequency/severity of claims for evaluating pure premium of each
policyholder under a rate--making system.

\begin{center}\scriptsize{\tiny
Table 1: Available covariates information for each  policyholder.
\begin{tabular}{c l}
  \hline
  Variable & Description \\
  \hline
  Gender & Equal to 0 for woman \& 1 for man\\
  Age & Equal to 1 for $18\leq age< 30;$ 2 for $30\leq age< 40;$ 3 for $40\leq age< 50;$ \&
  4 for $50\leq age$ \\
  Car's price & Equal to 1 for $price<2\times10^4;$ 2 for $2\times10^4\leq price<5\times10^4;$ 3 for $5\times10^4\leq
price<10^5;$ \& 4 for $10^5\leq price$\\
 Living area & Equal to 1 for
$population~size<10^5;$ 2 for $10^5\leq
population~size<5\times10^5;$ 3
for $5\times10^5\leq population~size<10^6;$\\
& \& 4 for $10^6\leq
population~size$\\
  \hline
\end{tabular}}\normalsize
\end{center}

For simplicity in presentation hereafter, we represent $kINBM_m$
for a k-Inflated Negative Binomial model with $m$ mixture
components and Pareto$M_m$ for a Pareto model with $m$ mixture
components.

To find an appropriate distribution for the frequency of claim, in
the first step, we considered the $kINBM_m$ model along with all
distributions that have been considered, by authors, to model
frequency of claims in a rate--making system. Namely, we
considered the kINBM, Delaporte, Sichel, and Poisson Inverse
Gaussian, say PIG, distributions for frequency and the Pareto$M_m$
distribution for severity and estimate their parameters.

The maximum likelihood estimator for the $kINBM_m$ we develop our
R codes while the maximum likelihood estimator for other
distributions have been computed using the GAMLSS package in R.
Table 2 represents the maximum likelihood estimator for
significant parameters of such distributions. The significant test
for each parameter has been tested by the Wald test.

Now using a backward elimination selection method, we find
covariates that may impact on response variable for each
regression model. The significant test for each covariate has been
done by the Wald test. Table 3 shows result of the backward
selection method for frequency/severity of accidents.

\begin{landscape}
\begin{center}\scriptsize{\tiny
Table 2. Estimation for parameters on various model for
frequency/severity
of claims.\\
\begin{tabular}[c]{lllllllll}
\hline\\
\multicolumn{7}{c}{{\bf Distribution:}} \\
   \hline\\
%\multirow{12}{0.001in}{\rotatebox{90}{\underline{{{\bf Parameters}}}}}  &
                          $NBM_1$                               & $0INBM_1$                         & $1INBM_1$                      & $2INBM_1$                 & $3INBM_1$                          & $NBM_2$                                 & $0INBM_2$  \\
   \hline\\
    ${\boldsymbol \omega}=(*,1)$ & ${\boldsymbol \omega}=(0.001,0.999)$ & ${\boldsymbol \omega}=(0.136,0.861)$ & ${\boldsymbol \omega}=(0,1)$ & ${\boldsymbol \omega}=(0.001,0.999)$ & ${\boldsymbol \omega}=(*,0.005,0.995)$ & ${\boldsymbol \omega}=(0.001,0.004,0.995)$\\
     ${\boldsymbol \tau}=23.390$ & ${\boldsymbol \tau}=22.256$ & ${\boldsymbol \tau}=1.755$ & ${\boldsymbol \tau}=19.408$ & ${\boldsymbol \tau}=32.333$ & ${\boldsymbol \tau}=(14.152,141.857)$ & ${\boldsymbol \tau}=(26.027,124.000)$ \\
     ${\boldsymbol \alpha}=5.717$ & ${\boldsymbol \alpha}=5.376$ & ${\boldsymbol \alpha}=0.217$ & ${\boldsymbol \alpha}=4.734$ & ${\boldsymbol \alpha}=7.730$ & ${\boldsymbol \alpha}=(39.02,32.53)$ & ${\boldsymbol \alpha}=(71.74,30.31)$ \\
    \hline\\
\multicolumn{7}{c}{{\bf Distribution:}} \\
    \hline\\
    $1INBM_2$ & $2INBM_2$ & $3INBM_2$ & $NBM_3$ & $0INBM_3$ & $1INBM_3$ & Delaporte\\
\hline\\
${\boldsymbol \omega}=(0.116,0.831,0.053)$ & ${\boldsymbol \omega}=(0,1, 0)$ & ${\boldsymbol \omega}=(0.001,0.997,0.002)$ &${\boldsymbol \omega}=(*,0.014, 0.982,0.004)$ & ${\boldsymbol \omega}=(0.003,0.007,0.986,0.004)$ & ${\boldsymbol \omega}=(0.125,0.644,0.033,0.198)$& $\lambda=0.243$\\
${\boldsymbol \tau}=(8.174,2.690)$& ${\boldsymbol \tau}=(19.408,30.250)$ & ${\boldsymbol \tau}=(25.316,42.478)$ &${\boldsymbol \tau}=(124.000,124.000,30.250)$ & ${\boldsymbol \tau}=(141.857,141.857,27.571)$ & ${\boldsymbol \tau}=(4.587,2.623,4.181)$& $\sigma=77.67$\\
${\boldsymbol \alpha}=(0.807,2.305)$& ${\boldsymbol\alpha}=(4.735,6.327)$ & ${\boldsymbol \alpha}=(6.054,8.88)$ & ${\boldsymbol \alpha}=(30.47,29.48,85.17)$ & ${\boldsymbol \alpha}=(33.70,32.58,77.67)$ & ${\boldsymbol \alpha}=(0.357,2.628,0.729)$& $\nu=0.913$\\
\hline\\
\multicolumn{7}{c}{{\bf Distribution:}} \\
    \hline\\
    Sichel & PIG & Pareto$M_1$ & Pareto$M_2$ & Pareto$M_3$ &\\
 \hline\\
       $\mu=0.242$ & $\mu=0.242$& ${\boldsymbol\rho}=1$ & ${\boldsymbol\rho}=(0.519, 0.481)$& ${\boldsymbol\rho}=(0.332,0.321, 0.347)$\\
       $\sigma=NS$  & $\sigma=0.225$&${\boldsymbol\alpha}=1.871$ &${\boldsymbol\alpha}=(1.871,1.871)$ &${\boldsymbol\alpha}=(1.871,1.873,1.873)$\\
      $\nu=-4.961$& ---& ${\boldsymbol\gamma}=16.44$& ${\boldsymbol\gamma}=(16.43,16.44)$ &${\boldsymbol\gamma}=(16.44,16.43,16.43)$\\
   \hline\\
\end{tabular}\\
where the first element in ${\boldsymbol \omega}$ stands for
weight of inflated part and we use $*$ whenever the distribution
is non-inflated distribution and $NS$ stands for not significant
at 5\% level. }\normalsize
\end{center}\newpage
\begin{center}\scriptsize{\tiny
Table 3. Regression coefficients for various model for
frequency/severity of claims.\\
\begin{tabular}[c]{llllllll}
\hline\\
\multicolumn{6}{c}{{\bf Regression model:}} \\
   \hline\\
 &                 $NBM_1$                   & $0INBM_1$                      & $1INBM_1$                      & $2INBM_1$                      & $3INBM_1$                      & $NBM_2$                               & $0INBM_2$  \\
   \hline\\
&    ${\boldsymbol \omega}=(*,1)$             & ${\boldsymbol \omega}=(0.041,0.959)$   & ${\boldsymbol \omega}=(0.111,0.889)$   & ${\boldsymbol \omega}=(0.001,0.999)$   & ${\boldsymbol \omega}=(0.002,0.998)$   & ${\boldsymbol \omega}=(*,0.893,0.107)$     & ${\boldsymbol \omega}=(0.047,0.025,0.928)$\\
&     ${\boldsymbol \alpha}=7.560$           & ${\boldsymbol \alpha}=22.56$   & ${\boldsymbol \alpha}=0.440$   & ${\boldsymbol \alpha}=7.950$   & ${\boldsymbol \alpha}=9.980$   & ${\boldsymbol \alpha}=(0.074,22.01)$  & ${\boldsymbol \alpha}=(0.061,18.71)$ \\
Intercept & ${\boldsymbol \beta_0}=-0.738$    & ${\boldsymbol \beta_0}=-0.698$ & ${\boldsymbol \beta_0}=-0.887$ & ${\boldsymbol \beta_0}=-0.744$ & ${\boldsymbol \beta_0}=-0.745$ & ${\boldsymbol \beta_0}=(NS,-0.766)$   & ${\boldsymbol \beta_0}=(NS,-0.719)$ \\
Gender & ${\boldsymbol \beta_1}=NS$          & ${\boldsymbol \beta_1}=NS$     & ${\boldsymbol \beta_1}=NS$     & ${\boldsymbol \beta_1}=NS$     & ${\boldsymbol \beta_1}=NS$     & ${\boldsymbol \beta_1}=(NS,NS)$       & ${\boldsymbol \beta_1}=(NS,NS)$ \\
Age&  ${\boldsymbol \beta_2}=-0.118$         & ${\boldsymbol \beta_2}=-0.118$ & ${\boldsymbol \beta_2}=-0.213$ & ${\boldsymbol \beta_2}=-0.121$ & ${\boldsymbol \beta_2}=-0.114$ & ${\boldsymbol \beta_2}=(NS,-0.116)$   & ${\boldsymbol \beta_2}=(NS,-0.117)$ \\
Car's price & ${\boldsymbol \beta_3}=-0.189$ & ${\boldsymbol \beta_3}=-0.189$ & ${\boldsymbol \beta_3}=-0.263$ & ${\boldsymbol \beta_3}=-0.190$ & ${\boldsymbol \beta_3}=-0.196$ & ${\boldsymbol \beta_3}=(0.709,0.175)$ & ${\boldsymbol \beta_3}=(0.587,0.180)$\\
Living area&  ${\boldsymbol \beta_4}=NS$     & ${\boldsymbol \beta_4}=NS$     & ${\boldsymbol \beta_4}=NS$     & ${\boldsymbol \beta_4}=NS$     & ${\boldsymbol \beta_4}=NS$     & ${\boldsymbol \beta_4}=(NS,NS)$       & ${\boldsymbol \beta_4}=(NS,NS)$ \\
    \hline\\
\multicolumn{6}{c}{{\bf Regression model:}} \\
    \hline\\
 &    $1INBM_2$                                    & $2INBM_2$                                & $3INBM_2$                            & $BM_3$                                     & $0INBM_3$                                   & $1INBM_3$ & Delaporte\\
\hline\\
&   ${\boldsymbol \omega}=(0.120,0.833,0.047)$            & ${\boldsymbol \omega}=(0.002,0.024,0.974)$     & ${\boldsymbol \omega}=(0.002,0.022,0.976)$ &${\boldsymbol \omega}=(*,0.93,0.02,0.05)$   & ${\boldsymbol \omega}=(0.03,0.07,0.46,0.44)$  & ${\boldsymbol \omega}=(0.13,0.07,0.40,0.40)$   & $\sigma=59.56$\\
&   ${\boldsymbol \alpha}=(0.685,9.471)$            & ${\boldsymbol\alpha}=(0.087,19.00)$      & ${\boldsymbol \alpha}=(0.090,18.94)$ & ${\boldsymbol \alpha}=(31.22,30.95,0.030)$ & ${\boldsymbol \alpha}=(13.29,11.32,0.157)$  & ${\boldsymbol \alpha}=(9.562,0.643,0.396)$   & $nu=0.910$\\
Intercept & ${\boldsymbol \beta_0}=(1.009,0.663)$    & ${\boldsymbol \beta_0}=(NS,-0.775)$      & ${\boldsymbol \beta_0}=(NS,-0.775)$  & ${\boldsymbol \beta_0}=(.0.638,1.032,NS)$  & ${\boldsymbol \beta_0}=(0.531,0.384,0.165)$ & ${\boldsymbol \beta_0}=(0.535,2.514,2.754)$  & ${\boldsymbol \beta_0}=-0.749$ \\
Gender&  ${\boldsymbol \beta_1}=(-0.185,NS)$        & ${\boldsymbol \beta_1}=(NS,NS)$          & ${\boldsymbol \beta_1}=(NS,NS)$      & ${\boldsymbol \beta_1}=(NS,NS,NS)$         & ${\boldsymbol \beta_1}=(NS,NS,NS)$          & ${\boldsymbol \beta_1}=(0.770,0.632,NS)$     & ${\boldsymbol \beta_1}=NS$ \\
Age&  ${\boldsymbol \beta_2}=(NS,-0.442)$           & ${\boldsymbol \beta_2}=(NS,-0.115)$      & ${\boldsymbol \beta_2}=(NS,-0.114)$  & ${\boldsymbol \beta_2}=(0.305,0.337,NS)$   & ${\boldsymbol \beta_2}=(NS,0.151,0.165)$    & ${\boldsymbol \beta_2}=(0.343,0.548,0.813)$  & ${\boldsymbol \beta_2}=-0.113$ \\
Car's price&  ${\boldsymbol \beta_3}=(0.607,0.076)$ & ${\boldsymbol \beta_3}=(0.618,0.177)$    & ${\boldsymbol \beta_3}=(0.675,0.179)$& ${\boldsymbol \beta_3}=(NS,0.676,0.332)$   & ${\boldsymbol \beta_3}=(0.230,0.151,0.310)$ & ${\boldsymbol \beta_3}=(NS,0.298,1.545)$     & ${\boldsymbol \beta_3}=-0.189$\\
Living area&  ${\boldsymbol \beta_4}=(NS,NS)$       & ${\boldsymbol \beta_4}=(NS,NS)$          & ${\boldsymbol \beta_4}=(NS,NS)$      & ${\boldsymbol \beta_4}=(NS,NS,NS)$         & ${\boldsymbol \beta_4}=(NS,NS,NS)$          & ${\boldsymbol \beta_4}=(0.238,2.382,0.623)$  & ${\boldsymbol \beta_4}=NS$ \\
\hline\\
\multicolumn{6}{c}{{\bf Regression model:}} \\
    \hline\\
 &    Sichel                          & PIG                                                        & Pareto$M_1$                    & Pareto$M_2$                             & Pareto$M_3$ &\\
 \hline\\
&                        $\sigma=NS$                        & $\sigma=0.174$                & ${\boldsymbol\rho}=1$       & ${\boldsymbol\rho}=(0.542, 0.458)$       & ${\boldsymbol\rho}=(0.341,0.312, 0.347)$\\
&                        $\nu=-5.688$                        & ---                            &${\boldsymbol\alpha}=1.927$     &${\boldsymbol\alpha}=(1.927,1.927)$      &${\boldsymbol\alpha}=(1.93,1.93,1.93)$\\
Intercept                & ${\boldsymbol \beta_0}=-0.737$ & ${\boldsymbol \beta_0}=-0.736$     & ${\boldsymbol \beta_0}=16.15$ & ${\boldsymbol \beta_0}=(6.15,6.15)$   & ${\boldsymbol \beta_0}=(16.15,16.15,16.15)$ \\
Gender                  & ${\boldsymbol \beta_1}=NS$     & ${\boldsymbol \beta_1}=NS$         & ${\boldsymbol \beta_1}=NS$     & ${\boldsymbol \beta_1}=(NS,NS)$         & ${\boldsymbol \beta_1}=(NS,NS,NS)$             \\
Age                     & ${\boldsymbol \beta_2}=-0.119$ & ${\boldsymbol \beta_2}=-0.119$     & ${\boldsymbol \beta_2}=NS$     & ${\boldsymbol \beta_2}=(NS,NS)$         & ${\boldsymbol \beta_2}=(NS,NS,NS)$             \\
Car's price             & ${\boldsymbol \beta_3}=-0.190$ & ${\boldsymbol \beta_3}=-0.190$     & ${\boldsymbol \beta_3}=0.157$  & ${\boldsymbol \beta_3}=(-0.16,-0.16)$ & ${\boldsymbol \beta_3}=(-0.16,-0.16,-0.16)$ \\
Living area             & ${\boldsymbol \beta_4}=NS$ & ${\boldsymbol \beta_4}=NS$             & ${\boldsymbol \beta_4}=NS$     & ${\boldsymbol \beta_4}=(NS,NS)$         & ${\boldsymbol \beta_4}=(NS,NS,NS)$             \\
   \hline\\
\end{tabular}\\
where the first element in ${\boldsymbol \omega}$ stands for
weight of inflated part and we use $*$ whenever the distribution
is non-inflated distribution and $NS$ stands for not significant
at 5\% level.}\normalsize
\end{center}\end{landscape}

\subsection{Model comparison}
To obtain an appropriate model for a given rate--making system,
this section begins by considering the $kINBM_m$ model along with
all distributions that have been considered, by authors, to model
frequency of claims in a rate--making system. Now in order to
compare result of regression/distribution models, we conducted
three evaluation approaches. Namely: ({\bf 1}) In the first
approach, to study performance of count distributions, we employ
each fitted distribution, 200 times, to simulate 8874 data. Then,
using the mean square error, say MSE, criteria, we compare
stimulated data with observed data (see Table 3 for result on such
comparison); ({\bf 2}) The second approach provides a pairwise
comparison between fitted count regression/distribution models
based upon {\it either } the Vuong test (for two non-nested
models) {\it or} the likelihood ratio test (for two nested
models), see Table 4 for such comparison study; and finally, ({\bf
3}) The third approach employs the Akaike Information Criterion
(AIC) and the Schwarz Bayesian information Criterion (SBIC) to
compare regression/distribution models for both frequency and
severity of claims, result of such comparison has been reported in
Table 5.

\subsection*{Generating Data approach:}
To study performance of fitted count distributions given in Table
1. We employ the GAMLSS package in R to generate samples from the
Delaporte, the Sichel, the Poisson Inverse Gaussian distributions.
Lim et al. (2014) introduced an idea to generate sample from a
given Zero-inflated Poisson mixture distribution. Now, we employ
their idea to generate samples from a given a $kINBM_m$
distribution. Based upon their idea, to generate sample $y_i$ from
a $kINBM_m$ distribution with probability mass function
\begin{eqnarray*} P(Y=y|{\boldsymbol \theta}) &=&
  p_1I_{\left(y=k\right)}+\sum_{j=2}^{m}p_j\left(\genfrac{}{}{0pt}{}{y+{\alpha }_j-1}{y}\right){\left(\frac{{\tau }_j}{{\alpha }_j+{\tau }_j}\right)}^{{\alpha }_j}{\left(\frac{{\alpha }_j}{{\alpha }_j+{\tau
  }_j}\right)}^y,
\end{eqnarray*}
where all parameters ${\boldsymbol\theta}=({\boldsymbol\alpha},
{\boldsymbol\tau}, {\boldsymbol p})$ and $k$ are given. We start
with a dummy variable, say $s_i,$ which generated from an uniform
(0,1) distribution. If $0\leq s_i\leq p_1,$ we set $y_i=k;$ If
$p_1< s_i\leq p_1+p_2,$ then $y_i$ is a draw from a Negative
Binomial distribution $(\alpha_1,\alpha_1/(\alpha_1+\tau_1));$ If
$p_1+p_2< s_i\leq p_1+p_2+p_3,$ then $y_i$ is a draw from a
Negative Binomial distribution $(\alpha_2,
\alpha_2/(\alpha_2+\tau_2));$ If $p_1+p_2+p_3< s_i\leq
p_1+p_2+p_3+p_4,$ then $y_i$ is a draw from a Negative Binomial
distribution $(\alpha_3, \alpha_3/(\alpha_3+\tau_3));$ and so on.

We employed the GAMLSS package and the above idea to simulate 8874
data (200 times). Table 4 reports mean (mean square error, say
MSE) of frequency for such 200 times simulated samples.

\begin{center}\scriptsize{\tiny
Table 4.  Mean and the MSE of frequency for generated data under count distributions given in Table 1.\\
\begin{tabular}[c]{ccccccc}
\hline\\
\multicolumn{5}{c}{{\bf Mean (MSE) of frequency for generated data under Distribution:}} \\
   \hline\\
Observed(Freq.) & Delaporte & Sichel & PIG & $NBM_1$ &$0INBM_1$&      $1INBM_1$ \\
   \hline\\
   0(6956) & 7018.240(5134.78)  & 7027.435(6384.40)& 7001.370(4548.36) & 69980.515(1579.14)& 7001.825(2964.21)& 6958.575(1498.88)\\
   1(1751) & 1614.615(19105.90) & 1584.62(28655.82)& 1620.145(22066.28)& 1626.805(13343.36)& 1625.580(16799.97)& 1748.300(1598.76)\\
   2(122)  & 203.210(6774.82)   & 227.235(11188.30)& 224.160(11090.70) & 222.960(11472.20) & 221.325(11189.41)& 120.835(136.98)\\
   3(31)   & 25.540(55.10)      & 29.680(26.90)    & 25.585(49.58)     & 23.645(81.46)     & 22.950(73.84)& 32.555(33.04)   \\
   4(9)    & 6.470(12.50)       & 4.230(24.28)     & 2.485(39.78)      & 1.910(48.10)      & 2.145(44.23) & 9.510(9.50)     \\
   5(3)    & 2.940(1.82)        & 0.595(6.12)      & 0.230(7.78)       & 0.175(6.94)       & 0.175(7.90) & 2.910(2.54)     \\
   6(2)    & 1.430(1.36)        & 0.160(3.74)      & 0.000(4.00)       & 0.000(4.00)       & 0.00(4.00)   & 0.895(1.54)     \\
   $>6$(0) & 1.560(2.88)        & 0.000(0.00)      & 0.000(0.00)       & 0.000(0.00)       & 0.00(0.00)    & 0.420(0.53)    \\
   \hline\\
\multicolumn{5}{c}{{\bf Mean (MSE) of frequency for generated data under Distribution:}} \\
   \hline\\
Observed(Freq.) &    $2INBM_1$ & $3INBM_1$ & $NBM_2$ & $0INBM_2$ & $1INBM_2$ & $2INBM_2$ \\
   \hline\\
  0(6956)    & 6988.835(2476.51) &7003.555(3652.21)  & 7019.265(6530.50) & 7019.135(3004.08)& 6958.730(1529.84)&7002.304(3033.16)\\
  1(1751)    & 1624.415(15079.51)&1626.055(16892.21) & 1619.010(22792.00)& 1616.830(6657.27)& 1748.690(1448.56)&1621.425(16415.15)\\
  2(122)     & 232.030(11339.35) &214.305(8663.16)   & 198.745(6347.56)  & 200.715(10290.01)& 122.380(124.82)  &220.400(10608.91)\\
  3(31)      & 26.405(72.11)     &28.205(20.45)      & 23.780(46.54)     & 23.940(35.47)& 30.530(24.16)    &26.855(53.16) \\
  4(9)       & 2.015(41.56)      &1.750(51.70)       & 7.010(11.76)      & 7.150(10.84) & 9.030(8.68)      &2.800(38.85)  \\
  5(3)       & 2.910(2.54)       &0.155(8.51)       &3.635(2.79)        & 3.350(3.60)   & 3.115(2.56)      &0.155(9.01)   \\
  6(2)       & 0.115(3.85)       &0.210(3.38)        & 1.730(1.18)       & 1.640(2.04) & 1.075(1.26)      &0.095(3.175)  \\
  $>6$(0)    & 0.000(0.00)       &0.000(0.00)        & 1.11(3.44)        & 1.090(2.13) & 0.450(0.60)      &0.000(0.000)\\
     \hline\\
\multicolumn{5}{c}{{\bf Mean (MSE) of frequency for generated data under Distribution:}} \\
   \hline\\
   Observed(Freq.)            & $3INBM_2$       &$BM_3$             &$0INBM_3$ & $1INBM_3$\\
   \hline\\
   0(6956) &7019.200(5955.36) &7015.995(1585.00)  &7019.785(7553.77)  &6959.035(1127.89)\\
   1(1751) &1605.105(22133.02)&1619.510(11018.88) &1617.120(21482.31) &1747.870(1038.81)\\
   2(122)  &215.755(7985.44)  &200.635(8149.00)   &199.745(5686.31)   &121.985(213.64)\\
   3(31)   &32.100(56.55)     &24.635(65.20)      &24.355(124.61)     &30.630(52.47)\\
   4(9)    &1.650(54.61)      &6.9100(10.76)      &6.955(12.64)       &9.605(8.34)\\
   5(3)    &0.205(9.04)       &3.635(2.78)        &3.400(3.20)        &3.245(1.94)\\
   6(2)    &0.000(4.00)       &1.600(1.14)        &1.595(0.675)       &1.215(1.37)\\
   $>6$(0) &0.000(0.00)       &0.650(1.28)        &1.045(2.01)        &0.415(0.68)\\
   \hline\\
\end{tabular}
}\normalsize
\end{center}

Results of the simulation study, given in Table 4, shows that the
MSE for the all 1-Inflated Negative Binomial mixture
distributions, is considerably less than the MSE of other fitted
distributions. Therefore, based upon this simulation study, one
may conclude that the 1-Inflated Negative Binomial mixture
distributions are appropriate distributions for claim frequency of
Iranian policyholders.

\subsection*{The Vuong and the likelihood ratio tests' approach:}
To make a decision about statistical hypothesis
\begin{eqnarray*}
H_0 &:& ~\hbox{Observed data came from a population with distribution function}~F\\
H_1 &:& ~\hbox{Observed data came from a population with
distribution function}~G.
\end{eqnarray*}
If both of distributions are belong to a family of distributions
with different parameters (nested models), one may employ the
likelihood ratio test to make such decision. Otherwise, where
models are belong to two different family of distributions
(Non-nested models) the Vuong has to used, see Denuit et al.
(2007, \S 2) for more details.

Table 5 represents a pairwise comparison between fitted count
regression/distribution models given in Tables 1 and 2.
\begin{center}\scriptsize{\tiny
Table 5.  Result of the Vuong test (for two non-nested models)
{\it or} the likelihood ratio test (for two nested
models).\\
\begin{tabular}[c]{cccc}
\hline\\
\multicolumn{4}{c}{{\bf Panel A:} Result of the Vuong test} \\
   \hline\\
Model 1 & Model 2 & Decision on fitted regression& Decision on fitted distribution  \\
   \hline\\
Delaporte  & $1INBM_1$ & $1INBM_1$ (Statistic=-37.63 \& P-value=0.00)& $1INBM_1$ (Statistic=-51.58 \& P-value=0.00)\\
PIG        & $1INBM_1$ & $1INBM_1$ (Statistic=-33.15 \& P-value=0.00)& $1INBM_1$ (Statistic=-38.94 \& P-value=0.00)\\
$0INBM_1$ & $1INBM_1$ & $1INBM_1$ (Statistic=-24.75 \& P-value=0.00)& $1INBM_1$ (Statistic=-35.57 \& P-value=0.00)\\
$2INBM_1$ & $1INBM_1$ & $1INBM_1$ (Statistic=-26.80 \& P-value=0.00)& $1INBM_1$ (Statistic=-37.09 \& P-value=0.00)\\
$3INBM_1$ & $1INBM_1$ & $1INBM_1$ (Statistic=-22.41 \& P-value=0.00)& $1INBM_1$ (Statistic=-31.22 \& P-value=0.00)\\
$0INBM_2$ & $1INBM_2$  & $1INBM_2$ (Statistic=-43.51 \& P-value=0.00)& $1INBM_2$ (Statistic=-52.27 \& P-value=0.00)\\
$2INBM_2$ & $1INBM_1$  & $1INBM_2$ (Statistic=-40.89 \& P-value=0.00)& $1INBM_2$ (Statistic=-37.22 \& P-value=0.00)\\
$3INBM_2$ & $1INBM_2$  & $1INBM_2$ (Statistic=-40.36 \& P-value=0.00)& $1INBM_2$ (Statistic=-33.16 \& P-value=0.00)\\
$0INBM_3$ & $1INBM_3$  & $1INBM_3$ (Statistic=-34.28 \& P-value=0.00)& $1INBM_3$ (Statistic=-20.48 \& P-value=0.00)\\
      \hline\\
\multicolumn{4}{c}{{\bf Panel B:} Result of the likelihood ratio test} \\
    \hline\\
Model 1 & Model 2 & Decision on fitted regression& Decision on fitted distribution  \\
   \hline\\
$NBM_1$  & $1INBM_1$ & $1INBM_1$    (Statistic=62.21 \& P-value=0.00)& $1INBM_1$ (Statistic=105.00 \& P-value=0.00)\\
$NBM_2$  & $1INBM_2$ & $1INBM_2$    (Statistic=80.46 \& P-value=0.00)& $1INBM_2$ (Statistic=49.66 \& P-value=0.00)\\
$NBM_3$  & $1INBM_3$ & $1INBM_3$    (Statistic=111.65 \& P-value=0.00)& $1INBM_3$ (Statistic=49.75 \& P-value=0.00)\\
$1INBM_1$  & $1INBM_2$ & $1INBM_2$ (Statistic=45.35 \& P-value=0.00)& $1INBM_1$ (Statistic=0.18 \& P-value=0.90)\\
$1INBM_1$  & $1INBM_3$ & $1INBM_2$ (Statistic=88.65 \& P-value=0.00)& $1INBM_1$ (Statistic=0.21 \& P-value=0.87)\\
$1INBM_2$  & $1INBM_3$ & $1INBM_3$ (Statistic=42.31 \& P-value=0.00)& $1INBM_2$ (Statistic=0.3 \& P-value=0.97)\\
\hline\\
\end{tabular}\\
}\normalsize
\end{center}

Based upon results of Table 5, one may conclude that the
1-Inflated Negative Binomial mixture distributions/regressions, at
5\% significant level, defeat other distributions/regressions.

\subsection*{The Akaike Information Criterion (AIC) and the
Schwarz Bayesian information criterion approaches:} The Akaike
Information Criterion (AIC) and the Schwarz Bayesian Information
Criterion (SBIC) are two measure to select an appropriate model
among a set of candidate models. Both criteria are defined based
on -2 times the maximum log-likelihood, penalized by {\it either}
number of estimated parameters, for AIC, {\it or} number of
estimated parameters times logarithm of number of observations,
for SBIC. Given a set of candidate models, a preferred model is
the one which has the minimum AIC (SBIC) value, see Denuit et al.
(2007, \S 1) for more details.

Table 6 provides the AIC and the SBIC for fitted
regression/distribution models for both frequency and severity of
claims.
\begin{center}\scriptsize{\tiny
Table 6.  Result of the Akaike Information Criterion
(AIC) and the Schwarz Bayesian information Criterion (SBIC).\\
\begin{tabular}[c]{ccccccc}
\hline\\
& & {\bf Regression model} && &{\bf Distribution model}&\\
\hline\\
Model & df & AIC & SBIC& df & AIC & SBIC  \\
   \hline\\
$NBM_1$     &6  &10656.42  &10698.88  &2  &10784.70  &10798.88\\
Delaporte   &7  &10615.42  &10665.07  &3  &10734.99  &10756.26\\
Sichel      &7  &10648.96  &10699.16  &3  &10772.67  &10793.94\\
PIG         &6  &10653.90  &10696.47  &2  &10781.11  &10795.29\\
$0INBM_1$   &7  &10664.59  &10714.25  &3  &10786.74  &10808.01\\
$1INBM_1$   &7  &10596.11  &10645.77  &3  &10681.69  &10702.96\\
$2INBM_1$   &7  &10658.36  &10708.02  &3  &10787.05  &10808.33\\
$3INBM_1$   &7  &10653.08  &10702.73  &3  &10783.25  &10804.53\\
$INBM_2$    &13 &10635.22  &10727.86  &5  &10735.14  &10770.59\\
$0INBM_2$   &14 &1064.07   &10746.80  &6  &10737.30  &10779.84\\
$1INBM_2$   &14 &10558.76  &10658.58  &6  &10687.80  &10730.02\\
$2INBM_2$   &14 &10638.66  &10748.39  &6  &10793.05  &10835.60\\
$3INBM_2$   &14 &10632.58  &10732.31  &6  &10789.88  &10832.42\\
$NBM_3$     &20 &10632.11  &10775.11  &8  &10741.26  &10797.99\\
$0INBM_3$   &21 &10654.22  &10803.49  &9  &10743.23  &10807.05\\
$1INBM_3$   &21 &10536.45  &10685.28  &9  &10693.19  &10757.33\\
Pareto$M_1$ &6  &63948.20  &63990.75  &2  &64102.30  &64116.48\\
Pareto$M_2$ &13 & 63968.20 &64054.38  &5  &64108.30  &64143.75\\
Pareto$M_3$ &20 &63974.20  &64108.93  &8  &64114.30  &64171.00\\
      \hline\\
\end{tabular}\\
}\normalsize
\end{center}

The AIC and SBIC for fitted models, given Table 6, show that the
1-Inflated Negative Binomial mixture distributions/regressions are
better than other distribution/regression models.
\section{Rate--making Examples}
To show practical application of our findings. We calculate the
rate and pure premiums for the set of well fitted
distributions/regression models that were presented in above
sections. Since we are interested in the differences between rate
premium of various classes. Therefore, we set the rate premium for
a new policyholder equal to 1 unite, at $t=0.$ Moreover, we
considered three different categories, described in Table 7.

\begin{center}\scriptsize{\tiny
Table 7: Categories which considered to evaluate rate and pure
premiums under well fitted models.
\begin{tabular}{c l}
  \hline
  Category & Description \\
  \hline
  $A_1$ & For a situation that no covariate information have been used for premium calculation\\
  $A_2$ & Whenever, chosen policyholder is a young man at age of 25 years old who owns a car\\ & with
price greater than $2\times 10^4$ and living in a city with population size larger than $10^6.$\\
  $A_3$ & Whenever, chosen policyholder is a mature woman at the age of 55 years old who owns\\ & a car with
price less than $2\times 10^4$ and living in a city with population size less than $10^5.$\\
  \hline
\end{tabular}}\normalsize
\end{center}
Now to calculate rate premium for three categories $A_1,$ $A_2,$
and $A_3,$ given in Table 7, using well fitted models. We consider
two different approaches. The first approach just considers number
of cumulated claims in the last yeas. While the second approach
considers the exact number of reported claim for each year in a
history of the policyholder. \footnote{It worthwhile to mention
that the second approach can be used just for inflated models.}

Tables 8 and 9 represent calculated rate premium for three
categories, given in Table 7, using well fitted models for both
approaches.

\begin{center}\scriptsize{\tiny
Table 8: The rate premium for three categories $A_1,$ $A_2,$ and
$A_3$ using well fitted models, whenever number of cumulated
claims has been considered.\\
\begin{tabular}{cc c cccccccc}
  \hline
\multicolumn{11}{c}{~~~~~~~~~{\bf Model:}}\\   \cline{3-11}\\
   & Number of cumulated &  &$NBM_1$  &  & & $NBM_2$  &  &  & $NBM_3$  &    \\
   \cline{3-11}\\
Year &claims up to this year ($K$) & $A_1$ & $A_2$ & $A_3$ & $A_1$ & $A_2$ & $A_3$ & $A_1$ &$A_2$  & $A_3$   \\
  \hline
$t=0$   & ---     & 1  & 1  & 1  & 1  & 1  & 1  & 1  & 1 & 1 \\ \hline\\
        & $K=0$  & 0.96 &  0.96 &   0.98 &   0.95 &  0.91  &  0.98   &   0.95 &  0.87  &  0.90 \\
        & $K=1$  & 1.13 & 1.08 & 1.11   & 1.02 & 1.03 & 1.10   &   1.02 & 1.00 &  1.09   \\
$t=1$   & $K=2$  &  1.29 & 1.21 & 1.24   &  1.54 & 1.18 & 1.35   &   1.42 & 1.13 & 1.96  \\
        & $K=3$  &  1.46 &  1.33 &  1.37  &  5.05 &  1.99  & 2.91   &   4.30 &  1.67 &  10.05  \\
        & $K=4$  &  1.63 & 1.46  & 1.50  &   10.40 & 5.78  & 9.12   &   9.76 & 4.88 &  24.99  \\ \hline\\
        & $K=0$  &  0.92 &  0.92 &   0.96  &  0.93  & 0.88  &  0.96   &  0.94 &  0.83  &  0.88   \\
        & $K=1$  &  1.08 & 1.04 & 1.09   &  0.97 & 1.00 & 1.08   &  0.98 & 0.97 & 1.04    \\
$t=2$   & $K=2$  &  1.24 & 1.16 & 1.22  &  1.04 & 1.06 & 1.19  &   1.04 &  1.05 & 1.24     \\
        & $K=3$  &  1.40 &  1.28 &  1.35   &  1.53 &  1.18 &  1.66   &   1.40  & 1.15 &  2.39  \\
        & $K=4$  &  1.57 & 1.40  & 1.47   &  4.69 & 1.60  & 4.00   &   3.92 & 1.41  & 8.56      \\
  \hline
\end{tabular}}\normalsize
\end{center}

\begin{center}\scriptsize{\tiny
Table 9: The rate premium for three categories $A_1,$ $A_2,$ and
$A_3$ using well fitted models, whenever exact number of reported
claim for each year of the policyholder's experience
has been considered.\\
\begin{tabular}{cc c cccccccc}
  \hline
\multicolumn{11}{c}{{\bf Model:}}\\   \cline{3-11}\\
  Year & Number of reported &  &$1INBM_1$  &  & & $1INBM_2$  &  &  & $1INBM_3$  &    \\
    \cline{3-11}\\
&claims at year $l$ ($k_l$) & $A_1$ & $A_2$ & $A_3$ & $A_1$ & $A_2$ & $A_3$ & $A_1$ &$A_2$  & $A_3$   \\
  \hline
$t=0$ & ---              &1      &1      &1   &1      &1      &1   &1      &1      &1     \\\hline\\
      & $k_1=0$         &0.64   &0.63   &0.88   &0.83   &0.81   &0.96   &0.79   &0.63   &0.96     \\
      & $k_1=1$         &1.81   &1.55   &1.54   &1.26   &1.30   &1.13   &1.39   &1.29   &1.17     \\
$t=1$ & $k_1=2$         &6.52   &3.91   &4.86   &2.50   &2.43   &2.55   &3.55   &2.54   &2.21      \\
      & $k_1=3$         &9.44   &4.94   &6.86   &3.30   &3.29   &3.64   &4.93   &3.50   &2.85      \\
      & $k_1=4$         &12.37  &6.38   &8.85   &4.13   &4.12   &4.54   &6.31   &4.46   &3.49      \\\hline\\
      & $k_1=0, k_2=0$  &0.48   &0.46   &0.78   & 0.72  &0.68   &0.93   &0.65   &0.48   &0.93      \\
      & $k_1=0, k_2=1$  &1.15   &1.03   &1.33   &1.06   &1.05   &1.09   &1.10   &0.84   &1.03      \\
      & $k_1=0, k_2=2$  &4.78   &2.56   &4.33   &2.19   &2.01   &2.46   &2.98   &1.79   &2.16      \\
      & $k_1=1, k_2=0$  &1.15   &1.03   &1.33   &1.06   &1.05   &1.09   &1.10   &0.84   &1.13      \\
$t=2$ & $k_1=1, k_2=1$  &2.87   &2.06   &2.31   &1.57   &1.61   &1.30   &1.90   &1.53   &1.16     \\
      & $k_1=1, k_2=2$  &6.79   &3.58   &5.63   &2.76   &2.60   &2.82   &3.93   &2.47   &2.38     \\
      & $k_1=2, k_2=0$  &4.78   &2.56   &4.33   &2.19   &2.01   &2.46   &2.98   &1.79   &2.15     \\
      & $k_1=2, k_2=1$  &6.79   &3.58   &5.63   &2.76   &2.60   &2.82   &3.93   &2.47   &2.38     \\
      & $k_1=2, k_2=2$  &9.10   &4.67   &7.88   &3.67   &3.34   &3.99   &5.31   &3.09   &3.37     \\
  \hline
\end{tabular}}\normalsize
\end{center}

To illustrate a guideline to use result of Tables 8 and 9, suppose
that {\it either} Negative Binomial with 2 mixture components,
$NBM_2,$ {\it or} 1-Inflated Negative Binomial with 2 mixture
components, $1INBM_2,$ can be considered as an appropriate model.
Now consider the following three different scenarios.
\begin{description}
    \item[Scenario 1:] For a given policyholder, no covariates information is available, category $A_1$ in Table
    7.  Based upon Table 8's and Table 9's result, respectively, his/her second year rate premium under $NBM_2$ model is 0.95 units
    while his/her second year rate premium under $1INBM_2$ model is 0.83 units, whenever such policyholder does not report any claim in the first
    year. But in the situation that such policyholder reports 2 claims in the
    first year. He/she has to pay 1.54 units, under $NBM_2$ model, and 2.50
    units, under $1INBM_2$ model.
    \item[Scenario 2:] The given policyholder belongs to category $A_2$ of Table
    7. Based upon Table 8's and Table 9's result, respectively, his second year rate premium, under $NBM_2$ model, is 0.91 units
    while his second year rate premium, under $1INBM_2$ model, is 0.81
    units, whenever such policyholder does not report any claim in the
    first year. But in the situation that such policyholder reports 2 claims in
    the first year. He has to pay 1.18 units, under $NBM_2$ model, and
    2.43 units, under $1INBM_2$ model.
    \item[Scenario 3:] The given policyholder belongs to category $A_3$ of Table 7. Based
upon Table 8's and Table 9's result, respectively, her second year
rate premium, under $NBM_2$ model, is 0.98 units while her second
year rate premium, under $1INBM_2$ model, is 0.96 units, whenever
such policyholder does not report any claim in the first year. But
in the situation that such policyholder reports 2 claims in the
first year. She has to pay 1.35 units, under $NBM_2$ model, and
2.55 units, under $1INBM_2$ model.
\end{description}
The above simple example, as well as other possible examples,
shows that: ({\bf 1}) the inflated models and covariates
information improve fairness of calculated rate premium; and ({\bf
2}) in the situation that number of reported claims uniformly
distributed in past experience of a policyholder (for instance
$k_1=1$ and $k_2=1$ instead of $k_1=0$ and $k_2=2$). His/Her rate
premium under inflated models is more fair and acceptable.

Now, to estimate the pure premium, we consider one mixture Pareto
distribution/regression model, as an appropriate model for claim's
severity, along with other well fitted counting models. Moreover,
we study situation that total claim size is {\it either} 1000
units (Case A) {\it or} 5000 unites (Case B). Table 10 and Table
11 show the pure premium under these assumptions.
\begin{landscape}
\begin{center}\scriptsize{\tiny
Table 10: The pure premium for three categories $A_1,$ $A_2,$ and
$A_3$ using well fitted models, whenever total claim size either
1000 or 5000 unites and exact number of reported claim for each
year of the policyholder's experience
has been considered.\\
\begin{tabular}{cc c cccccccc}
  \hline\\
  \multicolumn{11}{c}{{\bf Case A: Total of reported claim reach to 1000 unites}}\\
  \hline\\
\multicolumn{11}{c}{{\bf Model:}}\\   \cline{3-11}\\
   & Number of cumulated &  &$NBM_1$ \& Pareto$M_1$  &  & & $NBM_2$ \& Pareto$M_1$  &  &  & $NBM_3$ \& Pareto$M_1$  &    \\
    \cline{3-11}\\
Year &claims up to this year ($K$) & $A_1$ & $A_2$ & $A_3$ & $A_1$ & $A_2$ & $A_3$ & $A_1$ &$A_2$  & $A_3$   \\
  \hline
$t=0$ & ---     &   613.739 &723.848 &461.546    &   607.554 &730.992& 463.354    &  623.147 & 726.489 &449.722  \\\hline\\
      & $K=0$  &   589.189 &694.894 &452.315    &   574.376 &667.716 &453.895      &  594.470 &634.635& 405.854 \\
      & $K=1$  &   629.269& 723.241 &460.357    &   561.898 &650.758 &441.088      &  576.836 &631.235& 423.590 \\
$t=1$ & $K=2$  &   622.519& 707.323 &448.918    &   735.309 &652.794 &470.045      &  697.730 &620.015& 661.912 \\
      & $K=3$  &   621.617& 689.808 &440.058    &   2128.105&974.677& 900.347     &  1860.741& 811.248& 3013.693  \\
      & $K=4$  &   620.905& 680.503 &432.993    &   3921.243&2548.529& 2535.831     & 3775.850& 2137.017& 6735.508 \\\hline\\
      & $K=0$  &   564.640& 665.940 &443.084    &   568.061 &641.339 &446.372      &   587.939 &605.724 &396.865  \\
      & $K=1$  &   601.425& 696.455 &452.063    &   533.528 &626.073 &429.213      &    553.127 &607.359 &399.204 \\
$t=2$ & $K=2$  &   598.391& 678.095 &441.677    &   497.293 &580.094 &413.935      &    510.093 &570.899 &416.165  \\
      & $K=3$  &   596.071& 663.875 &433.633    &   643.233 &573.089 &510.020    &    603.755 &555.112 &714.869  \\
      & $K=4$  &   598.049& 652.537 &424.333    &   1769.746&702.016 &1106.136     &   1514.878& 613.404& 2298.398 \\
  \hline
  \multicolumn{11}{c}{{\bf Case B: Total of reported claim reach to 5000 unites}}\\
  \hline\\
\multicolumn{11}{c}{{\bf Model:}}\\   \cline{3-11}\\
   & Number of cumulated &  &$NBM_1$ \& Pareto$M_1$  &  & & $NBM_2$ \& Pareto$M_1$  &  &  & $NBM_3$ \& Pareto$M_1$  &    \\
    \cline{3-11}\\
Year &claims up to this year ($K$) & $A_1$ & $A_2$ & $A_3$ & $A_1$ & $A_2$ & $A_3$ & $A_1$ &$A_2$  & $A_3$   \\
  \hline
$t=0$ & ---    &  613.739& 723.848& 461.546   & 607.554& 730.992& 463.354   &   623.147 &726.489 &449.722\\\hline\\
      & $K=0$  &  589.189& 694.894& 452.315   & 574.376& 667.716& 453.895   &    594.470& 634.635& 405.854\\
      & $K=1$  &  799.370& 944.429& 550.863   & 713.788& 686.966& 456.491   &   732.764 &666.356 &438.382  \\
$t=1$ & $K=2$  &  790.796& 923.642& 537.174   & 934.074& 689.115& 486.460   &   886.338 &654.512 &685.028\\
      & $K=3$  &  789.650& 900.770& 526.572   & 2703.365& 1028.907& 931.789 &   2363.729& 856.385& 3118.939\\
      & $K=4$  &  788.745& 888.620& 518.119   & 4981.215& 2690.327& 2624.388&   4796.520& 2255.920& 6970.728\\\hline\\
      & $K=0$  &  564.640& 665.940& 443.084   & 568.061 &641.339  & 446.372 &   587.939 &605.724  &396.865 \\
      & $K=1$  &  763.999& 909.450& 540.937   & 677.749 &643.612  &436.741  &    702.646& 624.374 &406.205\\
$t=2$ & $K=2$  &  760.145& 885.475& 528.510   & 631.719 &596.345  &421.194  &    647.979 &586.893 &423.463\\
      & $K=3$  &  757.198& 866.907& 518.885   & 817.108 &589.144  &518.964  &    766.960 &570.663 &727.407\\
      & $K=4$  &  759.711& 852.102& 507.757   & 2248.135 &721.683 &1125.535 &    1924.372 &630.588 &2338.707\\
  \hline
\end{tabular}}\normalsize
\end{center}
\end{landscape}

\begin{landscape}
\begin{center}\scriptsize{\tiny
Table 11: The pure premium for three categories $A_1,$ $A_2,$ and
$A_3$ using well fitted models, whenever total claim size either
1000 or 5000 unites and exact number of reported claim for each
year of the policyholder's experience
has been considered.\\
\begin{tabular}{cc c cccccccc}
  \hline\\
  \multicolumn{11}{c}{{\bf Case A: Total of reported claim reach to 1000 unites}}\\
  \hline\\
\multicolumn{11}{c}{{\bf Model:}}\\   \cline{3-11}\\
  Year & Number of reported &  &$1INBM_1$ \& Pareto$M_1$  &  & & $1INBM_2$ \& Pareto$M_1$  &  &  & $1INBM_3$ \& Pareto$M_1$  &    \\
    \cline{3-11}\\
&claims at year $l$ ($k_l$) & $A_1$ & $A_2$ & $A_3$ & $A_1$ & $A_2$ & $A_3$ & $A_1$ &$A_2$  & $A_3$   \\
  \hline
$t=0$ & --         &     623.391 & 757.191&  571.461&   611.317 &    733.685 &     534.525&  609.443&  790.408&     566.132\\\hline\\
      & $k_1=0$    &     398.970 & 477.030  &   502.886 & 507.393 &   594.285  &   513.144 & 481.452 & 497.957 & 543.487\\
      & $k_1=1$    &     1023.795 & 1085.798 & 790.796     & 698.893  &  882.399   &  542.755 & 768.625 & 943.307  & 595.197\\
$t=1$ & $k_1=2$    &     3195.859 & 2390.932 &  2178.476  & 1201.672 &   1439.796  & 1069.149 &  1701.115 & 1621.325 & 981.386\\
      & $k_1=3$    &     4019.222 & 2562.144 &  2203.502 &1405.025 &1706.367 &1169.205 &2099.022 &1815.284 &915.449      \\
      & $k_1=4$    &      4712.021 &2973.705 &  2554.659 &1573.213 &1920.324 &1310.525 &2403.626 &2078.797 &1007.430    \\\hline\\
      & $k_1=0, k_2=0$ & 219.378 & 259.780 & 332.448 & 322.694  &  372.101  &  366.774 & 290.423  &   282.966 & 388.461\\
      & $k_1=0, k_2=1$ & 650.477 & 721.531 & 682.960 & 587.958 & 712.707 & 523.542 & 608.265 & 614.247 & 523.977\\
      & $k_1=0, k_2=2$ & 2342.976 & 1565.418 & 1940.906 & 1052.665 & 1190.942 & 1031.414 & 1427.979 & 1142.587  & 954.742\\
      & $k_1=1, k_2=0$ & 650.477 & 721.531 & 682.960 & 587.958 & 712.707 & 523.542 & 608.265  & 614.247 & 523.977\\
$t=2$ & $k_1=1, k_2=1$ & 1406.766 & 1259.673  & 1035.449 & 754.650 & 953.939 & 545.056 & 910.456 & 976.625 & 559.523\\
      & $k_1=1, k_2=2$  &  2936.409 & 1942.306 & 2239.077 & 1170.474 & 1366.822 & 1049.038 & 1661.517 & 1398.870 & 937.710\\
      & $k_1=2, k_2=0$  &  2342.976 & 1565.418 & 1940.906 & 1052.665 & 1190.942 & 1031.414 & 1427.979 & 1142.587 & 954.742\\
      & $k_1=2, k_2=1$  &   2936.409  & 1942.306 & 2239.077 & 1170.474 & 1366.822 & 1049.038 & 1661.517 & 1398.870 &  937.710\\
      & $k_1=2, k_2=2$  &   3520.914 & 2276.944  & 2816.357 & 1392.471 & 1577.924 & 1333.877 & 2008.510 & 1572.678 & 1193.225\\
  \hline
  \multicolumn{11}{c}{{\bf Case B: Total of reported claim reach to 5000 unites}}\\
  \hline\\
\multicolumn{11}{c}{{\bf Model:}}\\   \cline{3-11}\\
  Year & Number of reported &  &$1INBM_1$ \& Pareto$M_1$  &  & & $1INBM_2$ \& Pareto$M_1$  &  &  & $1INBM_3$ \& Pareto$M_1$  &    \\
    \cline{3-11}\\
&claims at year $l$ ($k_l$) & $A_1$ & $A_2$ & $A_3$ & $A_1$ & $A_2$ & $A_3$ & $A_1$ &$A_2$  & $A_3$   \\
  \hline
$t=0$ & --              &   623.391 &757.191 &571.461 &611.317 &   733.685 &   534.525 &609.443 &790.408 &566.132\\\hline\\
      & $k_1=0$         &   398.970 &    477.030 &502.886 &507.393 &594.285 &513.144 &481.452 &497.957 &543.487\\
$t=1$ & $k_1=1$         &   1300.542 &  1417.866 &946.265 &887.815 &1152.261 &649.459 &976.396 &  1231.797 &712.211\\
      & $k_1=2$         &   4059.749 &3122.146 &2606.761 &1526.503 &1880.126 &1279.341 &2160.953 &2117.171 &1174.325\\
      & $k_1=3$         &   5105.682 &3345.717 &2636.704 &1784.825 &2228.221 &1399.067 &2666.421 &2370.447 &1095.424\\
      & $k_1=4$         &   5985.752 &3883.148 &3056.902 &1998.477 &2507.613 &1568.174 &3053.363 &2714.552 &1205.490      \\\hline\\
      & $k_1=0, k_2=0$  &   219.378 &259.780 &332.448 &322.694 &   372.101 &366.774 &290.423 &282.966 &388.461\\
      & $k_1=0, k_2=1$  &   826.311 &942.195 &817.229 &746.892 &930.673 &626.469 &772.688 &802.100 &626.989\\
      & $k_1=0, k_2=2$  &   2976.319 &2044.167 &2322.484 &1337.216 &1555.166 &1234.188 &1813.983 &1492.022 &1142.443\\
      & $k_1=1, k_2=0$  &   826.311 &   942.195 &817.229 &746.892 &930.673 &   626.469 &772.688 &802.100 &626.989\\
$t=2$ & $k_1=1, k_2=1$  &   178.7037 & 1644.916 &1239.016 &958.644 &1245.680 &652.213 &1156.567 &  1275.304 &669.525\\
      & $k_1=1, k_2=2$  &   3730.166 &2536.317 &2679.275 &1486.871 &1784.835 &1255.277&2110.651&1826.684 &1122.062\\
      & $k_1=2, k_2=0$  &   2976.319 &2044.167 &  2322.484 &1337.216 & 1555.166 &1234.188 &1813.983 &1492.022 &1142.443\\
      & $k_1=2, k_2=1$  &   3730.166 &253.6317 &  2679.275 &148.6871 &1784.835 & 1255.277&2110.651&1826.684 &1122.062\\
      & $k_1=2, k_2=2$  &   4472.672 &2973.297 &3370.048 &1768.877 &2060.497 &1596.115 &2551.441 &2053.647 &1427.811\\
  \hline
\end{tabular}}\normalsize
\end{center}
\end{landscape}
Same as the above, to illustrate a guideline to use result of
Tables 10 and 11, suppose that {\it either} Negative Binomial with
2 mixture components, $NBM_2,$ {\it or} 1-Inflated Negative
Binomial with 2 mixture components, $1INBM_2,$ can be considered
as an appropriate model for claim frequency. Now consider the
following three different scenarios.
\begin{description}
    \item[Scenario 1:] For a given policyholder in category $A_1$ of Table
    7.  Based upon Table 10's and Table 11's result, respectively, his/her second year pure premium under $NBM_2$ model is 622.519 units
    while his/her second year pure premium under $1INBM_2$ model is 3195.859 units, whenever such policyholder reported 2
    claims with total size 1000 units in the first
    year. But in the situation that total size of two reported claims reach to 5000 units. He/she has to pay 790.796 units, under $NBM_2$ model, and
    4059.749 units, under $1INBM_2$ model.
    \item[Scenario 2:] The given policyholder belongs to category $A_2$ of Table
    7. Based upon Table 10's and Table 11's result, respectively, his second year pure premium, under $NBM_2$ model, is 707.323 units
    while his second year pure premium, under $1INBM_2$ model, is 2390.932
    units, whenever such policyholder reported 2 claims with total size 1000 units in the first
    year. But in the situation that total size of two reported claims reach to 5000 units. He has to pay 932.642 units, under $NBM_2$ model, and
    3122.146 units, under $1INBM_2$ model.
    \item[Scenario 3:] The given policyholder belongs to category
$A_3$ of Table 7. Based upon Table 10's and Table 11's result,
respectively, her second year pure premium, under $NBM_2$ model,
is 440.918 units while her second year pure premium, under
$1INBM_2$ model, is 2178.476 units, whenever such policyholder
reported 2 claims with total size 1000 units in the first year.
But in the situation that total size of two reported claims reach
to 5000 units. She has to pay 537.174 units, under $NBM_2$ model,
and 2606.761 units, under $1INBM_2$ model.
\end{description}
The above simple example shows that: ({\bf 1}) the inflated models
provides more fair pure premium of policyholders who made some
claims in their past experience. While for both cases A and B, the
pure premium under non-inflated models do not fairly penalized
such policyholders; and ({\bf 2}) in the situation that number of
reported claims uniformly distributed in past experience of a
policyholder (for instance $k_1=1$ and $k_2=1$ instead of $k_1=0$
and $k_2=2$). His/Her pure premium under inflated models is more
appealing and acceptable.
\section{Conclusion and suggestion}
This article introduces an k-Inflated Negative Binomial mixture
(kIBNM) distribution/regression model and provides an EM algorithm
to estimate its parameters. As an application of the kIBNM
distribution/regression to model number of reported claim under a
rate--making system has been given. Moreover, in order to compute
the pure premium under the system, severity of reported claim has
been model with a Pareto mixture distribution/regression model. As
an application frequency of reported claim of Iranian third party
liability, at 2011, has been model by the kIBNM and all of
possible models that have been used by authors. Numerical
illustration shows that: ({\bf 1}) the kIBNM models provide more
fair rate/pure premiums for policyholders under a rate--making
system; and ({\bf 2}) in the situation that number of reported
claims uniformly distributed in past experience of a policyholder
(for instance $k_1=1$ and $k_2=1$ instead of $k_1=0$ and $k_2=2$).
The rate/pure premium under the kIBNM models are more appealing
and acceptable.

We conjecture that the result of this article may be improved by
considering a Double Inflated Negative Binomial with probability
mass function $P(Y=y|{\boldsymbol
\theta})=p_1I_{k_1}(y)+p_2I_{k_2}(y)+\sum_{j=3}^{m}p_j\left(\genfrac{}{}{0pt}{}{y+{\alpha
}_j-1}{y}\right){\left(\frac{{\tau }_j}{{\alpha }_j+{\tau
}_j}\right)}^{{\alpha }_j}{\left(\frac{{\alpha }_j}{{\alpha
}_j+{\tau
  }_j}\right)}^yI_{{\Bbb
N}}(y),$ where $k_1,k_2\in{\Bbb N},$ $\sum_{j=1}^{m}p_j=1,$ and
$p_j,\alpha_j,\tau_j\geq0,$ for all $j=1,\cdots,m.$

%\section*{Acknowledgements}
%Authors would like to thank professor Jean-Philippe Boucher and
%professor Jean Lemaire  for their constructive suggestions which
%improved results and presentation of this article.

\end{document}